# Orbit, meteoroid size, and cosmic ray exposure history of the Aguas Zarcas CM2 breccia

Peter JENNISKENS[1,2*], Gerardo J. SOTO[3], Gabriel GONCALVES SILVA[4], Oscar LÜCKE[3], Pilar MADRIGAL[3], Tatiana BALLESTERO[3], Carolina SALAS MATAMOROS[5], Paulo RUIZ CUBILLO[5,6], Daniela CARDOZO MOURAO[7], Othon CABO WINTER[7], Rafael SFAIR[7,8], Clemens E. TILLIER[9], Jim ALBERS[1,9], Laurence A. J. GARVIE[10], Karen ZIEGLER[11], Qing-Zhu YIN[12], Matthew E. SANBORN[12], Henner BUSEMANN[13], My E. I. RIEBE[13], Kees C. WELTEN[14], Marc W. CAFFEE[15,16], Matthias LAUBENSTEIN[17], Darrel K. ROBERTSON[2], and David NESVORNY[18]

[1] SETI Institute, 339 Bernardo Ave, Mountain View, CA 94043, USA.

[2] NASA Ames Research Center, Moffett Field, CA 94035, USA.

[3] Escuela Centroamericana de Geología, Universidad de Costa Rica, San José, 2060, Costa Rica.

[4] Institute of Chemistry, University of Sao Paulo, Sao Paulo, SP 05508-000, Brasil.

[5] Centro de Investigaciones Espaciales, Escuela de Fisica, Universidad de Costa Rica, San José, 2060, Costa Rica.

[6] Labaoratorio Nacional de Materiales y Modelos Estructurales, LANAMME.

[7] Grupo de Dinamica Orbital e Planetologia, Sao Paulo State University - UNESP, Guaratinguetá, CEP 12516-410 Sao Paulo, Brasil.

[8] Laboratory for Instrumentation and Research in Astrophysics, Paris Observatory, Paris, France.

[9] Lockheed Martin, Advanced Technology Center, Palo Alto, CA 94304, USA.

[10] Center for Meteorite Studies, School of Earth & Space Exploration, Arizona State Univ., Tempe, AZ 85287, USA.

[11] Institute of Meteoritics, University of New Mexico, Albuquerque, NM 87131, USA.

[12] Department of Earth and Planetary Sciences, University of California at Davis, Davis, CA 95616, USA.

[13] Institute of Geochemistry and Petrology, ETH Zürich, CH-8092 Zürich, Switzerland.

[14] Space Sciences Laboratory, University of California, Berkeley, CA 94720, USA.

[15] Department of Physics and Astronomy/PRIME Laboratory, Purdue University, West Lafayette, IN 47907, USA.

[16] Department of Earth, Atmospheric and Planetary Sciences, Purdue University, West Lafayette, IN 47907, USA.

[17] Gran Sasso National Laboratory, National Institute for Nuclear Physics, I-67100 Assergi, Italy.

[18] SWRI, Department of Space Studies, Boulder, CO 80302, USA.

*Corresponding author. Email: Petrus.M.Jenniskens@nasa.gov

**Abstract –** The Aguas Zarcas (Costa Rica) CM2 carbonaceous chondrite fell during night time in April 2019. Security and dashboard camera video of the meteor were analyzed to provide a trajectory, lightcurve, and orbit of the meteoroid. The trajectory was near vertical, 81° steep, arriving from an ~109° (WNW) direction with apparent entry speed of 14.6 ± 0.6 km/s. The meteoroid penetrated to ~25 km altitude (5 MPa dynamic pressure), where the surviving mass shattered, producing a flare that was detected by the Geostationary Lightning Mappers on GOES-16 and GOES-17. The cosmogenic radionuclides were analyzed in three recovered meteorites by either gamma-ray spectroscopy or accelerator mass spectrometry (AMS), while noble gas concentrations and isotopic compositions were measured in the same fragment that was analyzed by AMS. From this, the pre-atmospheric size of the meteoroid and its cosmic-ray exposure age were determined. The studied samples came from a few cm up to 30 cm deep in an object with an original diameter of ~60 cm, that was ejected from its parent body 2.0 ± 0.2 Ma ago. The ejected material had an argon retention age of 2.9 Ga. The object was delivered most likely by the 3:1 or 5:2 mean motion resonances and, without subsequent fragmentation, approached Earth from a low $i < 2.8°$ inclined orbit with perihelion distance $q = 0.98$ AU close to Earth orbit. The steep entry trajectory and high strength resulted in deep penetration in the atmosphere and a relatively large fraction of surviving mass.

## INTRODUCTION

Aguas Zarcas is a CM2 chondrite breccia that fell in Costa Rica at 21:07:22–26 local time on the night of 2019 April 23, corresponding to 03:07:22–26 UTC on April 24 (Soto *et al.,* 2019a,b). This is the fourth CM-type carbonaceous chondrite fall to be documented by videos or photographs of the meteor. The fall follows Sutter's Mill (Jenniskens *et al.,* 2012), Maribo (Haack *et al.,* 2013; Borovicka *et al.,* 2019), and Flensburg (Borovicka *et al.,* 2021), potentially providing a new clue about the source of CM chondrites in the asteroid belt (Dunn *et al.,* 2013; Jenniskens, 2014; Borovicka *et al.,* 2015; Granvik & Brown, 2018). Since the fall of Aguas Zarcas, Winchcombe added another CM2 chondrite to this record (King et al., 2022).

Aguas Zarcas is the most significant mass of CM2 type meteorites to fall since Murchison (~100 kg – Fuchs et al., 1973; Pizzarello et al., 2020). The recovered mass was at least ~27 kg (Lücke *et al.,* 2019; Soto *et al.,* 2019b; Kichinka, 2019; Kappler, 2019) and may have been as high as 30 kg. The first reported fragment had a mass of 1,152 g and pierced through the corrugated zinc roof overhang of a house in La Caporal de Aguas Zarcas, split a cable and a wooden support beam, and landed between two collapsible plastic tables that were stored there vertically. The stone was still warm when it was pulled from in between the two tables a minute or so later. Further fragments of <1 g to 1,875 g were recovered in the following days northwest of Aguas Zarcas, in the villages of La Cocaleca, Santa Rosa, and La Palmera.

Before Aguas Zarcas, only one other meteorite had been recovered in Costa Rica, in the city of Heredia following a witnessed fall on April 1, 1857. A tiny part of that H5 chondrite meteorite is preserved at the Central American School of Geology of the University of Costa Rica (Soto, 2004).

A cosmochemical characterization of Aguas Zarcas was recently published by Kerraouch *et al.* (2022), concluding that Aguas Zarcas was a brecciated CM chondrite with distinct lithologies, including two distinct metal-rich lithologies, a CM1/2 lithology, a C1 lithology, and a brecciated CM2 lithology consisting of different petrologic types. These different lithologies were formed under different conditions and experienced different levels of hydration. The breccia contains strongly shocked fragments redistributed in unshocked lithology (Yang et al., 2022). Hydration played a role in the evolution of the organic matter in Aguas Zarcas (Dionnet *et al.,* 2022). Organic compounds are diverse (Pizzarello *et al.,* 2020; Aponte *et al*., 2020), but already showed signs of terrestial contamination (Tunney *et al.,* 2022). Ti and Cr isotopes were studied by Torrano *et al.* (2021).

Due to the large and relatively pristine amount of material available, Aguas Zarcas has been used in impact experiments (Michikiami *et al.,* 2023) and outgassing studies (Yu *et al.,* 2021). Infrared reflection spectroscopy was used for comparisson to possible parent asteroids (Fieber-Beyer *et al.,* 2021). Thermophysical properties were measured by Opeil et al. (2024).

Unfortunately, Aguas Zarcas did not fall in a dedicated camera network (e.g., Borovicka *et al.,* 2019, 2021; Devillepoix *et al.,* 2020). It necessitated an analysis of lesser quality video security and dash-cam data. One video with two stars was obtained by monitoring cameras of the National Seismological Network (University of Costa Rica) at the top of the Turrialba Volcano (Lücke *et al.,* 2019). The meteor was filmed also by video security cameras at Quepos, Heredia and San José and by a dashboard camera in Heredia and San Ramón (Fig. 1). A preliminary orbit was derived based on the Quepos and Turrialba Volcano videos alone by a team from the Brazilian Meteors Observation Network and the University of Sao Paulo (Soto *et al.,* 2019a; Takir *et al.,* 2020).

Here, we re-visit the available video observations to determine the trajectory and orbit of the meteor, as well as the manner of energy deposition in the atmosphere. The meteorite strewn field is described. Some of the recovered meteorites were analysed to determine the meteoroid's original size and its cosmic-ray exposure (CRE) history. The results are discussed in the context of other documented falls with CM2 affinity.

## EXPERIMENTAL METHODS

### Trajectory and orbit

Trajectory and orbit were derived from video security camera and dash cam footage distributed via social media channels (Table 1a). The camera locations were derived from Google Earth to better than 2-m accuracy. In all videos, the meteor moved downwards at a small angle from vertical, even though the directions from meteor to camera span a wide range: from Az = 122° from North (Turrialba Volcano) to Az = 203° (San Ramón) (Fig. 1).

The footage was analyzed using methods in Jenniskens *et al.* (2012) and Popova *et al.* (2013), but in this case no star background calibration images were made at the camera sites. Instead, we used stars in the video from Turrialba Vulcano to calibrate this line of sight, and the foreground features in other videos to determine their image projection and scale. That left relatively large systematic errors for the placement of the horizon in the latter. We then triangulated the position of the end flare in latitude, longitude and altitude. We confirmed that this position was consistent with the location of recovered meteorites. Subsequently, we aligned the astrometry of each video to match the position of the flare and used the relative astrometry to that point to derive planes that fit the trajectory orientation and the meteor's velocity profile using the CAMS software suite (Jenniskens et al., 2011). By forcibly matching the lines of sight to the end flare, systematic errors in the position of the horizon are removed. In this way, the image scale per pixel and how well the meteor position is defined in pixels determines most of the uncertainty in instantaneous velocity, and the systematic and random errors can be deduced from combining different pairs of lines of sight.

Original videos taken from the observatories at the Turrialba and Poás Volcanos were made available in mp4 compression. Videos from security camreas and dashboard cameras from the towns of Turrialba, Heredia, Quepos and San José were downloaded from posted videos on youtube. The posted videos were captured, individual frames extracted, and the position of the leading fragment marked for each. The position of the horizon and cardinal directions were determined, after which a *Stellarium* star field in fisheye projection was overlaid to determine the azimuth and elevation of these positions in the field of view. The program GIMP v.2.10 was used to correct the images for lens barrel distortion (Huang *et al.,* 2016) based on the lens-induced curvature of straight objects in the video.

The low-light *Turrialba Volcano camera 2* video (Fig. 2A) was made with a Starlight Network IR Mini Bullet Camera (IPC2122SR3-UPF40 (60)-C) with a 1/2.9" progressive scan 2.0 migapixel CMOS sensor and 4.0mm F1.4 lens. It recorded 30 frames-per-second (fps) from about 600 m East of the volcano's crater using a lens with a wide-angle 85° horizontal field of view (fov) aimed

West (270°). Unique images were only provided as 12 frames per second. The location of the camera was verified and found mounted 8.21 m above the ground. The meteor is seen through thin clouds. The video has two recognizable stars in the unclouded part of the image: Sirius (alpha Canis Majoris) and Betelgeuse (alpha Orionis). This allowed for an absolute calibration of the position and orientation of the horizon (rotated by -0.0 ± 0.1°). Starfields on unclouded nights in the months prior were used to define the barrel distortion. The meteor entered the field of view just prior to the flare. The meteor moves straight down, parallel to constant azimuth lines.

The video from the village of Turrialba used is the steady black and white video (Fig. 2B), not the version filmed from screen. No stars are visible in the images. The location of the camera is uncertain by ~10 m, as it could be on either side of the street. The meteor is near the center of the image, entering at the top of the frame and then moving straight down. The meteor initially moves in and out of thin clouds, then in and out of thicker clouds, and the final part is obscured. Foreground features (buildings, a wall, and a series of streetlights at ~480 m distance) that were identified in Google Earth show the camera has a 91.2 ± 2.3° horizontal field of view aimed to Az = 316.5° ± 0.4° (NW). The rotation of the field of view (+2.0 ± 0.5°) and horizon location were calibrated from parallel lines in the building and the position of the streetlights. The line of distant lamps along a road perpendicular to the viewing direction gives the plate scale. The barrel distortion correction factor was -30 ± 5 in Gimp, based on the shape of the lamp pole seen right in the field of view.

The street intersection at 9.91497ºN, 83.68337ºW is 10 m higher than the camera station, which required a 1.2º correction. But the street is inclined 664 m left edge to 652m right edge, with 649m.

The *Heredia gate* video, from a security camera watching a gate (Fig. 2C), also shows an early part of the meteor, and then all the way down to where the final fragments fade. The meteor is seen in part through thin clouds. Visible above the gate are the contours of a hill or mountain. The view is towards the Poás Volcano. Parallel lines in the wall and position of distant lamps imply a rotation of the horizon by +5.6 ± 0.1°, assuming the bricks of the wall are installed on the level and the distant lamps are at the same elevation. The horizon was placed based on these distant lamps. The field of view is about 90° and a Gimp factor of about -60 removed the curvature of the lines in the foreground. The location of the camera in the city of Heredia is unknown, we adopted the geographic center of the city and adjusted the azimuth to match the position of the flare derived from other data (see below).

The Quepos video (Fig. 2D) was taken at 4 unique frames per second with a high-placed security camera near a building at the Marina Pez Vela. The view is over a bay, which helps define the orientation of the horizon (-5.1 ± 0.2°). The water level defines a constant elevation surface, with the camera positioned above that by 7–10m. The range to the edge of the Bay varies in different parts of the image. The view also includes foreground palm trees, a billboard and a sign, all of which define the azimuth directions. The camera has a 90° horizontal field of view and the center of the image was pointed 11.2 ± 0.5° E from N. Countours of the building help define the location of the horizon. Parallel lines in the parking lot define the barrel lens distortion GIMP factor as -30 ± 2. A small systematic error of 0.4° in azimuth was found during triangulation (Table 1a). The meteor entered the field of view during the plateau in brightness (see below). The flare itself was in between frames and not captured, but the color change marks the time of the event. Early on, the meteor had a blueish color, but the final flare was distinctly redder in color. The trajectory is tilted slight from left to right.

The San Ramón video (Fig. 2E) was obtained using a 60Hz 1080p resolution DashCam YI with an ultrawide 165° by 93° field of view from a driving car on a straight road heading 7.2° E from North. There are no stars visible in the partially clouded sky. The meteor entered the field of view just before the onset of the final flare. The perspective at that moment was used for calibration. The road and approximate location of the video was identified by the observer (given as 10.08099°N, 84.47019°W) and refined using Google Earth. Road center lines, edges, cables and buildings alongside the road provide the perspective that defines the position of the horizon. The video orientation and scale were calibrated using foreground objects, confirming the factory provided field of view (167.5 ± 2.5° horizontal field of view). The orientation of the field of view (-2.2 ± 0.3°) and location of the horizon were calibrated from distant palm trees and streetlight series perspectives. Note that the elevation of the road increased towards the intersection. The barrel distortion correction factor was -30 ± 5 in GIMP based on the shape of the car dashboard.

The San Ramón video did not change gain and no saturation correction was applied to the extracted photometry.

The video security camera from San José at the *Rotonda de la Bandera* (Fig. 2F) was located in the NW corner of the building of the Economic Science Guild in San Pedro of Montes de Oca. Only a low-resolution version of this video is available. The field of view is ultra wide-angle, about ~135° wide, and the horizon is highly curved and rotated (by -7.8 ± 0.1°), because the camera is looking down at an. angle, with flag poles defining the lens barrel distortion with GIMP factor -60 ± 5. The position of the horizon is well defined by lights in the far field. While distant features are present in the image, the azimuthal directions were calibrated on recognizable features around and just beyond the rotonda at the center of the image, with the scale adjusted to match the azimuth to the flare position.

Finally, a dashboard camera captured the meteor from Heredia (Table 1a), shortly after the car passed an unidentified church. The car was just exiting a shallow curve in the road. In collecting the meteor positions in subsequent frames, it was assumed that the horizon stayed constant during this turn. Meteor positions were aligned along the direction of the meteor wake (Fig. 3). The horizon was placed from perspective lines in the road. Lamp posts define the image rotation to +0.0 ± 0.1° and GIMP factor of about -30 ± 3. To put the meteor at about the right elevation and direction, the field of view of the camera was about 100° and the car moved in a direction 358° (North).

**Meteor lightcurve**

To derive the meteor lightcurve, we extracted sum-pixel intensities for each frame in the available videos (Table 1a) and subtracted the background signal derived from nearby positions. The intensities were corrected for saturation (when specified in Sect. 2), and for the range to the meteor, normalizing to a standard distance of 100 km. No elevation-dependent extinction correction was made, possibly affecting the shape of the lightcurve slightly. These lightcurves were aligned in magnitude (logarithmic scale) to trace the relative brightness of the meteor. The time-dependent brightness change during the flare was traced from the relative brightness profile of reflected light on flat surfaces.

The video-derived lightcurve traces were aligned with the Geostationary Lightning Mapper (GLM) detected brightness of the final flare, which provided both a calibrated time and a calibrated peak brightness. GLM onboard both the GOES-16 and GOES-17 weather satellites (Jenniskens *et al.,* 2018) detected the main flare just above the detection threshold, but the flare was not bright enough to be reported by the US Government Satellite network (Tagliaferri *et al.,* 1998).

**Meteorite samples**

Cosmogenic radionuclide activities were determined by means of non-destructive ultra-low background high purity germanium (HPGe) gamma-ray spectrometry on two small stones in the underground laboratory STELLA (SubTerranean Low-Level Assay) at the Laboratori Nazionali del Gran Sasso (LNGS) (Laubenstein, 2017; Arpesella, 1996). The 2.65 g stone #01 "Flight-oriented" was counted for 20.25 days starting 164 days after the fall and a 3.28 g stone #02 "Pre-rain" was counted for 34.07 days starting 185 days after the fall, respectively. The counting efficiencies for the detected radionuclides have been calculated using a thoroughly tested Monte Carlo code *G EANT4* software developed at CERN (Agostinelli *et al.,* 2003).

All other measurements reported here were performed on Arizona State University stone ASU#2121_5 (ASU identification number). Oxygen isotope studies were performed at the University of New Mexico on a number of samples, but the measurements reported here are for stone ASU#2121_5. The analytical methods for this procedure have been described previously (Popova *et al.,* 2013, Langbroek *et al.,* 2019).

Chromium isotope studies at the University of California, Davis were performed on three different stones labeled: ASU#2121_5, ASU#2121_6 (a chondrule-rich fragment), and ASU#2121_7 (from a chondrule-poor half-stone). The analytical methods for this procedure have been described previously also (Popova *et al.,* 2013; Langbroek *et al.,* 2019).

The ASU#2121_5 sample was prepared for cosmogenic radionuclide analysis at UC Berkeley. For the analysis of the long-lived cosmogenic radionuclides $^{10}$Be (half-life = $1.36 \times 10^6$ yr), $^{26}$Al ($7.05 \times 10^5$ yr), and $^{36}$Cl ($3.01 \times 10^5$ yr), we dissolved 104.1 mg in a mixture of concentrated $HF/HNO_3$, along with a carrier solution containing 3.2 mg of Be and 2.8 mg of Cl. After isolation of Cl as AgCl, a small aliquot - equivalent to 3.3% of the dissolved sample - was taken for chemical analysis by inductively coupled plasma optical emission spectroscopy (ICP-OES), and a second carrier solution containing 8.3 mg of Al was added to the remaining solution. Be, Al, and Cl were separated and purified using procedures described previously (e.g., Nishiizumi *et al.,* 2014). Concentrations of $^{10}$Be, $^{26}$Al, and $^{36}$Cl were measured by accelerator mass spectrometry (AMS) at Purdue University (Sharma *et al.,* 2000). The measured $^{10}$Be/Be, $^{26}$Al/Al, and $^{36}$Cl/Cl ratios were corrected for blank levels (which are <0.1% of the measured values) and normalized to AMS standards (Sharma *et al.,* 1990; Nishiizumi, 2004; Nishiizumi *et al.,* 2007) that were measured in the same run. The radionuclide concentrations (in atoms/g) were calculated by multiplying the measured radionuclide to stable nuclide ratios with the known amounts of stable Be, Cl and Al carrier added to the sample (taking into account the small amount of stable $^{27}$Al from the meteorite itself. The meteorite contains negligible amounts of stable Be and Cl). Concentrations are converted to activities in disintegrations per minute per kg of sample (dpm/kg) using the half-lives that were adopted for the AMS standards.

The CRE age was determined from noble gas concentrations in sample ASU#2121_5. Helium, Ne, Ar, Kr, and Xe concentrations and isotopic compositions were measured at ETH Zürich as described in Riebe *et al.* (2017). Two aliquots of the sample were analyzed, a larger aliquot of 28 mg and a smaller aliquot of 17 mg. The samples were degassed in a crucible at ~1700 °C. The larger aliquot was analyzed again at ~1750 °C to ensure that the samples were fully degassed, a so-called re-extraction. The re-extraction yielded <1.5% of the total gases in all cases, and in most isotopes significantly less. Blank corrections were in all cases ≤1%, except for $^{40}$Ar, where the blank correction was 4% on the large sample and 9% on the small sample.

## RESULTS

### Trajectory and orbit

All videos reliably associated with the fall of Aguas Zarcas show a nearly vertical descent of the bolide (Fig. 2). Seen from San Ramón, the meteor moved slightly from left to right. This is also the case from Quepos and from San José. Seen from Turrialba and Turrialba Volcano, the meteor moved straight down. This implies a steep entry angle and arriving from a WNW or NW azimuth.

Due to small errors in the placement of the horizon and image scale, each individual line of sight has systematic errors in the measured azimuth and elevation of the meteor. We constrained the solution by demanding that the end flare is a fixed marker of elevation and position. The median altitude of this flare measured by all stations is 25.1 ± 1.0 km (standard error). The results are summarized in Table 1b.

The triangulated solution of Turrialba Volcano 2, Turrialba, Quepos and San Ramón (using CAMS software described in Jenniskens et al., 2011) converged at 10.4091°N, 84.3695°W. The uncertainty in this position is about ± 1.0 km (± 0.0090º). This adopted position of the flare is consistent with the location of recovered meteorites on the ground (for wind drift, see below). We then adjusted the azimuth and elevation for the flare in all measured lines of sight for small systematic errors to match this position and altitude. Those errors range from 0.1° to 2.9° (Table 1a). As expected, the largest systematic errors are those from the ultra wide-angle dashboard camera video from San Ramón. However, that camera was also nearest to the meteor.

After triangulating the measured planes through meteor and station, the position of the flare based on *all* lines of sight came out to 10.4092°N, 84.3710°W. We find the meteor moved to azimuth 109.4° ± 5.9° (SE) with a steep elevation angle of 81.2 ± 1.6°. The uncertainty was evaluated by triangulating sets of data leaving one station out. The Right Ascension and Declination of the apparent radiant are R.A. = 165.8 ± 0.6° and +13.2 ± 1.0°. Soto et al. (2019b) reported an azimuth of 117° and elevation of 73°, with speed 14 km/s, based on the Turrialba Volcano and Quepos cameras alone (Table 2).

The entry speed was evaluated from the cameras at Turrialba and Heredia that contained the earliest parts of the trajectory. The meteor first entered the field of view of the Turrialba video at 42.8 ± 0.2 km altitude. Figure 4 shows the height versus time profiles measured for individual videos, based on the common solution of the trajectory. Small errors in scale, mostly for the Heredia cameras, resulted in different solutions for the velocity and deceleration. The post-flare section of the profile for the Heredia gate camera was fitted to those of other cameras to arrive at the result in Fig. 4. A least-squares fit to the early straight part of the slope, after taking into account the entry angle, results in initial velocities of 13.46 km/s for Heredia gate, 13.77 km/s for Turrialba, and 13.26 km/s for the Heredia dashboard camera.

These velocities were measured centered on the altitudes of 37.3, 37.8 and 36.4 km, respectively, relatively deep into the Earth's atmosphere but before the main fragmentations. Therefore, a small amount of pre-detection deceleration needs to be taken into account. The nominal change in velocity of a 60-cm sized meteoroid (see below) moving from 80 km to 43 km due to atmospheric friction is about 0.24 km/s, while that would have increased to 1.0 km/s at 35 km altitude. Indeed, significant decleration is observed only below 32 km altitude, after the onset of flares that are indicative of fragmentation (Fig. 5). When taking this pre-detection deceleration into account, the initial speed is 14.17 km/s for Heredia gate, 14.44 km/s for Turrialba and 14.06 km/s for Heredia dashboard camera.

It has been pointed out that assuming a constant speed at the beginning of the path can underestimate the initial velocity (Egal et al., 2017). The best approach to deriving the initial velocity is mutli-parameter fitting from simulated meteoroid decelerations (Gural, 2012). Alternatively, a linearly independent deceleration formulation can be developed (Sansom et al., 2019). The method by Gural (2012), as implemented in the CAMS software suite, results in a best-fit initial speed of 15.12 km/s. The uncertainty in this result is also dominated by systematic errors. Therefore, we combined this result with the previous velocity determinations and adopted an apparent entry speed of 14.6 ± 0.6 km/s (Table 2).

By the time the meteoroid shattered causing the final flare (Fig. 5), the speed had slowed down to 9.2 ± 0.4 km/s. In the Heredia dashcam video, the largest fragment was followed down to 18.2 km altitude, when the residual speed was only ~ 2–4 km/s.

**Satellite observations and absolute brightness**

The Geostationary Lightning Mappers on GOES-16 and GOES-17 detected the final flare of the meteor. The event triggered 19 Level 0 event detections on GOES-17, of which 12 are in Level 2 data, all in a single pixel. GOES-17 observed the meteor in the pixel projected on the lightning ellipsoid (located at 15.6 km altitude at this latitude – Jenniskens *et al.,* 2018) at 84.2097°W, 10.4149°N. The flare is centered at 24-Apr-2019 03:07:24.779 UTC. GOES-16 made a threshold detection in the pixel projected at 84.3942°W, 10.4591°N, 21-km further West-North-West. The event triggered 14 Level 0 events on GOES-16, but barely above threshold.

The GLM sensor on GOES-16 has the more top-down view (with GOES-16 in a 146° azimuth direction from the meteor). GOES-17 saw the flare from the side (with GOES-17 in a 270° azimuth direction). Triangulation weighted by event intensity results in an approximate position of the flare at 24 ± 4 km altitude over longitude ~ 84.37°W and latitude ~ 10.42°N ("GLM" in Fig. 6 below). The pixel size of GLM is 8 x 8 km. Within this uncertainty, the calculated position is in good agreement with the position of the flare derived from the video observations and the location of the meteorites, but the accuracy is insufficient to help constrain the location of the flare.

The reported intensity values were automatically corrected for the factor that is routinely applied to the OI line emission of lightning to account for the 777-nm OI line moving out of the filter passband in off-center field-of-view observations. Because the flare was closer to the edge of the field of view, the brightness of GOES-17 was corrected by a factor of 6.4 ± 0.2 upward, with the correction for GOES-16 being a factor of 1.15 ± 0.05 upward (Jenniskens *et al.,* 2018). Because the bolide emission at this wavelength is expected to be from continuum emission instead, the uncorrected brightness was used to calibrate the lightcurve. From this, and by assuming a 6000 K blackbody emission (Taggliaferri *et al.,* 1998; Jenniskens *et al.,* 2018), GLM on GOES-17 measured a total radiated energy $6.7 \times 10^7$ J and peak intensity $2.0 \times 10^8$ W/sr, while the GLM on GOES-16 measured a total radiated energy of $3.1 \times 10^7$ J and peak intensity $9.6 \times 10^7$ W/sr for the flare alone and integrated over wavelength. The GOES-16 derived intensity is a factor of two (0.75 magnitudes) lower, but the detection is near the threshold and less of the lightcurve was sampled. The peak brightness of the flare had an absolute magnitude $M_v$ = -13.82 ± 0.04 from the better defined lightcurve in GOES-17 data.

**Meteor lightcurve**

The video cameras provide a consistent light curve shape with a distinct flare near the end (Fig. 5), which was used to align the time scale and intensity scale to that of GLM. Some viewing directions from video cameras had thin clouds that became visible during the flare. Those lines of sight that did not show clouds provided a consistent lightcurve shape with a rapid rise, a weak initial flare and a long plateau, followed by a long flare and a brief intenser one.

Based on this entry speed and the recorded timing, the meteor lightcurve had an onset at ~70 km altitude. In comparison, the C1 chondrite Flensburg (19.5 km/s entry speed) was first detected at a height of 71.8 km in daytime conditions (Borovicka *et al.,* 2021). The overall shape of the lightcurve resembles that of the even faster Maribo (Borovicka & Spurny, 2019), in the sense that it rose early to a plateau. However, in the plateau many smaller flares occurred that are not seen in Aguas Zarcas (Fig. 5).

The integrated radiated energy along the dashed line contour is $4.1 \times 10^8$ J. If the luminous efficiency is the mean 0.0135% of carbonaceous chondrite (Type II) meteors (Revelle & Ceplecha, 2001), then the initial kinetic energy was ~$3.0 \times 10^{10}$ J, or ~0.0073 kt. This total energy is a factor of ten smaller than the lower limit of 0.073 kt for detections reported by the US Government satellites discussed in Taggliaferri *et al.* (1998).

From this kinetic energy and an entry speed of 14.6 km/s, the initial mass was approximately 280 kg. That corresponds to a meteoroid with diameter ~63 cm for a typical 2.2 g/cm$^3$ bulk density of carbonaceous chondrite meteorites (Macke et al., 2011; Opeil et al., 2020).

**Infrasound**

Infrasound of the Aguas Zarcas fall was detected by a nearby portable infrasound array I69CR (Sarapiqui, Costa Rica) at 10.44098°N, 84.02743°W, 62m and by a more distant station I20EC (Isla Santa Cruz, Galapagos Islands, Ecuador) part of the IMS-ICDC network (Villalobos-Villalobos & Quintero-Quintero, 2019). At station I69, the airburst was detected in 4 array elements in 0.1–8 Hz filtered data. The event time ranged from 03:08:29 to 03:10:59 UTC, with the signal passing with a speed of 388 m/s from a back azimuth direction 265.52 ± 0.02°. The frequency was f = 1.82 ± 0.18 Hz and peak amplitude 0.3 Pa.

At the 1390 km distant station I20EC (located on the Galapagos Islands, Ecuador, at 0.696°S, 90.309°W, 177m), the signal was detected by 8 elements H1, H2, H3 and H4 from a back azimuth

of 29.72 ± 0.01° at 04:22:12 to 04:32:33 UTC. The speed was 355 m/s and frequency 2.04 ± 0.88 Hz, with peak amplitude of 0.02 Pa.

The infrasound signal was also detected by a seismic network at the Turrialba volcano, where it arrived first from a 302.4°N back azimuth direction at 03:12:19 UTC with amplitude 0.13 Pa and apparent shock wave velocity of Cs = 347 m/s (Quintero-Quintero et al., 2019). The maximum aplitude of 0.23 Pa was reached at 03:12:23 UTC, arriving from 302.10 ± 0.05ºN and Cs = 362 m/s. At 03:12:34 UTC, the amplitude was 0.084 Pa and back azimuth 300.4ºN. The median apparent frequency at the station was f = 2.86 ± 1.58 Hz (Quintero-Quintero & Villalobos-Villalobos, 2019). The azimuth directions from Turrialba and I69CR triangulate to 10.413°N, 84.398°W, just west of the meteorite strewn field.

The average frequency of f = 2.04 ± 0.88 Hz would be produced by an initial bolide kinetic energy of about E = 0.0002–0.006 kt (nominally 0.00074 kt), using log (E) = -3.68 log (f) – 1.99 (Gi & Brown, 2017), or a slightly smaller 0.00015–0.0032 kt (nominally 0.00049 kt) using the more traditional log (E) = -3.34 log (f) – 2.28 (ReVelle, 1997). The blast radius R = Cs / (2.81 f), with Cs the sound speed at altitude, is only ~60 cm (in the range of 40–110 cm).

The close distance of station I69 to the event, and confirmation of the pressure amplitude from the seismic network, enables a check on this energy range from the measured apparent pressure amplitude. The value measured at the ground scales with the source altitude $\Delta p \sim (Po/P)^{-2/3}$, in which Po is the pressure at ground level and P is the pressure at the height of the airburst.

That kinetic energy corresponds to a meteoroid mass of 8–236 kg (nominally 29 kg) for an entry speed of 14.6 km/s. Given that as much as 27 kg of meteorites were recovered on the ground, the entry mass must have been in the higher range of this uncertainty interval. With a density of 2.2 $g/cm^3$, this corresponds to a likely diameter range of about 30 – 60 cm.

**Doppler weather radar signature and winds**

The Aguas Zarcas (Costa Rica, April 24, 2019, 03:07:24.8 UTC) fall was not detected by Doppler weather radars to our knowledge. The vertical wind profile showed mild 4–10 m/s winds towards various directions above 16 km altitude, shifting to moderately strong 13–27 m/s winds towards

W and NW between 16 and 9 km altitude, then shifting gradually back to mild 3–8 m/s winds towards East below 4 km altitude.

Because of the steep trajectory, the strewn field is expected to be compact. Falling from an altitude of 25.1 km/s with a residual speed of 9.2 km/s, 1-kg meteorites would have fallen 1.92 km E and 1.30 km S of the breakup point and impacted with a speed of 76 m/s (marked by open triangle in Fig. 6). 100-g meteorites would have fallen 1.70 km E and 1.55 km S, impacting at 52 m/s, while 10-g meteorites would have fallen 1.70 km E and 1.95 km S, impacting at 36 m/s. The predicted fall locations put 10-g meteorites only 0.7 km SSW from 1-kg meteorites (Fig. 6, triangles). These positions fall at the center of the meteorite strewn field (Fig. 6).

**Meteorite recovery**

Figure 7 shows one of the recovered meteorites. The larger recovered meteorites have highly irregular shapes and strong regmaglypts, indicative of strong ablation without much secondary fragmentation. Smaller stones are rounded.

Nearly all meteorites were found by local inhabitants of the area and purchased by meteorite dealers. Large masses and small masses were found mixed on the ground. Only a few have known fall locations (Table 3). The reported find locations cover an area 5.8 km long, along a line to azimuth 111°, and only 1.3 km wide (Fig. 6).

Rain hit the area starting on afternoon of April 27, 2019, lasting three days. It has been reported that in the pre-rain period of April 23–27 five meteorites of several hundred grams up to 1.8 kg were found, as well as several mid-sized 5–8 cm ones (with mass a few 100–200g). In addition, hundreds of fully intact specimens 1–4 cm in size were recovered. In later searches, more material was recovered up to an estimated mass of 27 kg (Soto *et al.,* 2019b).

**Meteorite mineralogy and petrography**

The mineralogy and petrography of Aguas Zarcas is presented at length in Kerraouch *et al.* (2020a; 2021), Kebukawa *et al.* (2020b) and Garvie (2021), and will not be repeated here. Aguas Zarcas, though officially classified as a brecciated CM2 in the Meteoritical Bulletin, is more complex. While dominated by CM lithologies, it also contains a significant quantity of unique carbonaceous chondrite lithologies. The identified lithologies include several CM1-2 materials varying in degree of aqueous alteration, a C1 lithology displaying neither chondrules nor CAI pseudomorphs, and two distinct types of unique metal-rich C2-3 lithologies which appear to be isotopically distinct from other carbonaceous chondrites. One of the metal-rich lithologies shows evidence of having been heated in excess of ~400°C.

**Oxygen and chromium isotopes**

The oxygen and chromium isotope studies serve to confirm that the Aguas Zarcas breccia belongs to the group of carbonaceous chondrites with CM affinity. Oxygen isotopes of Aguas Zarcas are listed in Table 4. As with other CM chondrites, results from individual fragments spread along the Carbonaceous Chondrite Anhydrous Mineral line in the CM domain, representing different levels of aqueous alteration (Clayton, 1993; Clayton *et al.,* 1991). Results for ASU#2121_5 further expand the range of known oxygen isotope values for Aguas Zarcas, having significantly lower Δ17O' than previously reported values measured in our lab (Soto *et al.,* 2019b).

The $\varepsilon^{54}$Cr – $\Delta^{17}$O' isotope diagram is shown in Fig. 8 and places Aguas Zarcas square among the other known CM type meteorites. Table 5 compares the $\varepsilon^{53}$Cr and $\varepsilon^{54}$Cr of Aguas Zarcas to other meteorites with CM affinity that have measured orbits. All show similar values.

**Noble gases**

The noble gases in Aguas Zarcas (Tables 6a–d) have compositions that are similar to noble gases in other primitive carbonaceous chondrites (Krietsch *et al.,* 2021; Mazor *et al.,* 1970). They are best understood as a mixture of primordial noble gases hosted in Phase Q and presolar grains, cosmogenic, and radiogenic noble gases. The $^{4}$He/$^{36}$Ar ratio of 46.63 ± 0.41 and 47.13 ± 0.44 and the $^{20}$Ne$_{tr}$/$^{36}$Ar ratios of 0.1846 ± 0.0014 and 0.1788 ± 0.0014 agree with a mixture of HL-bearing presolar nanodiamonds and Q (Busemann *et al.,* 2000; Huss & Lewis, 1994). The $^{36}$Ar/$^{132}$Xe (69.39 ± 0.56, 67.73 ± 0.56) and $^{84}$Kr/$^{132}$Xe (0.9104 ± 0.0077, 0.8755 ± 0.0078) ratios are in good agreement with ratios previously determined for Q (Busemann *et al.,* 2000; Wieler *et al.,* 1991, 1992).

The Ne three-isotope plot (Fig. 9) shows that Ne in Aguas Zarcas is a mix of primordial and cosmogenic noble gases. In order to determine cosmic ray exposure (CRE) ages, cosmogenic gases were therefore deconvolved from the trapped gases. In the Ne three-isotope plot, the two data points plot close to each other, but are not overlapping (Fig. 9). We estimate the range of the $^{20}Ne/^{22}Ne$ ratio in the trapped component ($^{20}Ne/^{22}Ne_{tr}$) by drawing two lines in the Ne three isotope plot, one from the upper end of the range of possible cosmogenic compositions to the large sample and one from the lower end of the possible cosmogenic compositions to the small sample. Extrapolating these lines to a typical trapped $^{21}Ne/^{22}Ne$ of 0.029 gives a $^{20}Ne/^{22}Ne_{tr}$ range of 7.0–7.8. We used these values together with a typical cosmogenic $^{21}Ne/^{22}Ne$ (($^{21}Ne/^{22}Ne)_{cos}$) range of 0.80–0.95 (Wieler, 2002) to determine concentrations of $^{21}Ne_{cos}$ and $^{20}Ne_{tr}$ (Table 6a, 7).

The $^{20}Ne/^{22}Ne$ ratio of the trapped gases of 7.0–7.8 is lower than the $^{20}Ne/^{22}Ne$ ratio of HL of 8.5 (Huss & Lewis, 1994). This indicates that Aguas Zarcas contains presolar SiC and/or graphite grains carrying Ne that is essentially pure $^{22}Ne$ ("Ne-E", Amari *et al.,* 1995; Lewis *et al.,* 1994).

The $^{3}He/^{4}He$ ratio also indicates a mix between primordial and cosmogenic He (Table 6a). The $^{3}He/^{4}He$ ratio of ~3 × $10^{-4}$ is higher than that of Q (1.23 × $10^{-4}$; Busemann *et al.,* 2000) and HL (≤1.70 × $10^{-4}$; Huss & Lewis, 1994), indicating that there is detectable $^{3}He_{cos}$ in the samples. The concentration of $^{3}He_{cos}$ was estimated assuming the trapped $^{3}He/^{4}He$ ratio to be in the range between Q and HL and the concentration of radiogenic $^{4}He$ ($^{4}He_{rad}$) to be between 0 and 2000 × $10^{-8}$ $cm^3/g$. The max $^{4}He_{rad}$ was determined based on 4.5 Gyr decay of U and Th of average CM chondrite concentrations (Lodders & Fegley, 1998).

The cosmogenic noble gases are not visible in In Ar, Kr, and Xe. Argon is most likely a mix of Q-Ar and radiogenic $^{40}Ar$. The $^{36}Ar/^{38}Ar$ ratios of both aliquots of 5.299 ± 0.025 and 5.271 ± 0.046 (Table 6b) are within uncertainty in agreement with the ratio previously determined for Q of 5.34 ± 0.02 (Busemann *et al.,* 2000). The Kr isotopic ratios are within 2σ uncertainties in agreement with Q (Table 6c). The Xe isotopic composition is in good agreement with Q with an addition of ~2% HL Xe from presolar diamonds (Fig. 10, Table 6d).

The production rates of cosmogenic $^{3}He_{cos}$ and $^{21}Ne_{cos}$ are dependent on the size of the meteoroid, the shielding of the sample in the meteoroid, and the elemental composition of the sample. The cosmogenic $^{22}Ne/^{21}Ne$ ratio can be used as a shielding indicator in samples with mostly cosmogenic Ne (Krietsch et al., 2021). Aguas Zarcas contains too much trapped Ne relative to cosmogenic Ne to do so. Based on the pre-atmospheric size of Aguas Zarcas from lightcurve (above) and cosmogenic nuclide data (see below), we determined maximum and minimum production rates for all shielding depths within a meteorite with a radius between 20 and 65 cm and average carbonaceous chondrite composition (Leya & Masarik, 2009). The outer 5 cm were excluded because the cosmogenic $^{10}Be$, $^{26}Al$ and $^{36}Cl$ concentrations indicate that our sample came from the interior portion of the meteoroid.

Resulting production rates as well as the CRE ages can be found in Table 7. The CRE age of Aguas Zarcas determined from the concentration of $^{21}Ne_{cos}$ and the adopted range of production rates is 1.7–2.7 Ma. The concentrations of $^{3}He_{cos}$ gave lower CRE ages of 0.4–0.5 Ma, indicating He-loss. Assuming that all $^{40}Ar$ is radiogenic and an average CM chondrite concentration of K of 370 ppm (Lodders & Fegley, 1998), and given that measured values were 230 ± 30 ppm for a few gram sample of Aguas Zarcas and about 530 ppm in a 0.1 g sample, we estimate the $^{40}Ar$ retention ages to 3.3 Ga for the large sample and 2.9 Ga for the small sample, indicating some Ar-loss since the formation of the solar system.

**Cosmogenic radionuclides from gamma-ray spectroscopy**

The two specimens measured by gamma-ray spectroscopy (Tables 8a, 8b) show different and low activities of $^{60}Co$ (#01: 32 ± 3 dpm/kg; #02: 2 ± 1 dpm/kg, in units of atom decays per minute and kg). This suggests that specimen #02 originated from <20 cm below the surface of the meteoroid with no significant production of secondary thermal neutrons, whereas the #01 specimen is from further inside.

The activities of the short-lived radioisotopes, with half-life less than the orbital period, represent the production integrated over the last segment of the orbit. The $^{22}Na/^{26}Al$ ratios of the two specimens are 1.26 ± 0.16 and 1.25 ± 0.14, respectively. The fall of Aguas Zarcas occurred during the end of the solar cycle 24 minimum, based on neutron monitor data (Bonino et al., 2001; Bartol, 2020). The cosmic ray flux was high in the six months before the fall. So the activities for the very short-lived radionuclides are expected to be high, as observed (Table 8a).

The naturally occurring radionuclides (Table 8b) uranium and thorium are high, but still in agreement with the average concentrations in CM chondrites (Braukmüller *et al.,* 2018). As shown in Nittler *et al.* (2004) and Braukmüller *et al.* (2018), the range of K concentrations in CM chondrites is quite wide and also covers values as low as the ones measured for this meteorite.

**Cosmogenic radionuclides from accelerator mass spectrometry**

The measured concentrations of most major elements in Aguas Zarcas (Table 9) are consistent with average CM chondrite composition (Wasson & Kallemeyn, 1988) and with the bulk composition of Aguas Zarcas reported previously (Kerraouch *et al.,* 2022). The K concentration in sample AZFM-2121-5 is 530 ppm, which is higher than the average CM chondrite value of 400 ppm, but again within the range for CM chondrites reported by Braukmüller *et al.* (2018).

The measured concentrations for major elements (in wt%) in sample AZFM-2121-5, minor elements (in ppm) and cosmogenic radionuclides (in dpm/kg) are shown in Table 9. The measured $^{10}$Be, $^{26}$Al and $^{36}$Cl concentrations of Aguas Zarcas are compared to calculated production rates for CM chondrites with radii of 20–150 cm (Leya & Masarik, 2009) to constrain its CRE history and pre-atmospheric size.

For meteorites with long CRE ages (>10 Ma), the radionuclide concentrations (in dpm/kg) are saturated and are equal to the production rate. The measured $^{10}$Be concentration of 15.5 ± 0.14 dpm/kg in Aguas Zarcas is 22–45% lower than the calculated $^{10}$Be production rates of 20–28 dpm/kg for CM chondrites with radii of 20-150 cm (Fig. 11a). The undersaturated $^{10}$Be concentration indicates a CRE age in the range of 1.5–3.0 Ma.

The measured $^{26}$Al concentration of 39.3 ± 0.9 dpm/kg overlaps with the lower end of calculated $^{26}$Al production rates of 32–56 dpm/kg for CM chondrites with radii of 20–150 cm (Fig 11b). Therefore, the $^{26}$Al concentration alone does not constrain the CRE age of Aguas Zarcas. However, when $^{10}$Be and $^{26}$Al are combined (Fig. 11c), the measured concentrations yield a best fit with the calculated production rates for a CRE age of 2.0 Ma, in good agreement with noble gas data.

## DISCUSSION

**Meteoroid size**

The corresponding $^{10}Be$ and $^{26}Al$ saturation values of ~24.2 and ~45.7 dpm/kg for Aguas Zarcas either indicate irradiation near the center of an object with R~30 cm, assuming an average density of 2.2 g/cm$^3$ for CM chondrites, or at 10–15 cm depth in an object with R = 40–100 cm.

Based on the CRE age inferred from $^{10}Be$ and $^{26}Al$, the $^{36}Cl$ concentration in Aguas Zarcas has reached 99% of the saturation value. The measured $^{36}Cl$ concentration of 60.5 ± 1.2 dpm/kg is a factor of 7–10 higher than calculated production rates of 6–8 dpm/kg from spallation reactions on K, Ca, Ti, Mn, Fe and Ni (Leya & Masarik, 2009), with Ca and Fe as the main target elements. The high $^{36}Cl$ indicates that Aguas Zarcas contains a large contribution of neutron-capture produced $^{36}Cl$ (54 ± 2 dpm/kg). If we assume an average CM chondrite composition with 160 ppm Cl (Wasson & Kallemeyn, 1988), this value corresponds to a $^{36}Cl$ production rate of ~340 dpm/gCl, comparable to the maximum production rate of ~300 dpm/gCl predicted for CI chondrites with radii of 50–100 cm gCl (Kollar *et al.,* 2006). If the bulk Cl content of Aguas Zarcas is closer to the average CM chondrite values of 400–500 ppm recently reported by Ebihara & Sekimoto (2019), then the high $^{36}Cl$ found in Aguas Zarcas corresponds to 110–135 dpm/gCl. This value is similar to the highest values found in the large Allende CV chondrite (Nishiizumi *et al.,* 1991). However, the neutron-capture depth profiles are also sensitive to the bulk H content of the meteorites, which is much higher in CM and CI chondrites than in CV chondrites, so a direct comparison between Aguas Zarcas and Allende is not possible. According to the model calculations of Kollar *et al.* (2006), values of 110–135 dpm/gCl are also found in the center of CI chondrite meteoroids as small as 25–30 cm. Since both H and Cl concentrations vary significantly among CM and CI chondrites, it is difficult to constrain the size of Aguas Zarcas from the $^{36}Cl$ result without knowledge of the H and Cl content of Aguas Zarcas.

However, despite these uncertainties, we conclude that the ASU sample of Aguas Zarcas (from a larger meteorite) came from a much more shielded position in the meteoroid than the two small samples measured by gamma-ray spectroscopy, which showed relatively low neutron-capture $^{60}Co$ activities of 2 and 32 dpm/kg, respectively, which are on the low end of the range of 14–226 dpm/kg found in Allende (Cressy 1972).

Constraints set by the lightcurve brightness (diameter ~ 63 cm) and by the infrasound frequency (diameter ~ 30–60 cm), suggest that the measured samples came from the center of the original meteoroid with diameter ~60 cm. For fragments near the center of the meteoroid to survive, the point of failure (Jenniskens et al., 2022) that caused the final flare was reached when only about half the meteorite had ablated. The initial mass was about 250 kg. With about 27 kg of material recovered, that implies that ≥11% of material survived to the ground, a relatively high fraction, representing up to 22% of the final part that penetrated intact to 25.1 km and fragmented. The amount of recovered material is likely a lower limit to the mass that survived, given that the strewn field contains the Aguas Zarcas river in the middle of the strewn field and a forested area in the northern part.

**Meteoroid strength**

The lightcurve of Aguas Zarcas is smoother than that of Maribo (Borovicka & Spurny, 2019), suggesting this was not such a weak meteoroid as Maribo. Indeed, Dionnet et al. (2022) found a low bulk porosity of 4.5 ± 0.5 vol% for Aguas Zarcas, compared to 18.3 – 24.9% for Murchison (Ciceri et al., 2021), but distributed inhomogeneously in the material with more pore space near chondrules.

The final flare was likely the moment where the residual mass no longer was able to withstand the pressure in front of the meteoroid (Jenniskens et al., 2022). Figure 12 shows the anticipated burst height as a function of yield strength of the Aguas Zarcas meteoroid (Robertson & Mathias, 2019). The meteoroid flare at 25.1 km altitude implies that it broke when the dynamical pressure was ~5 MPa, either measuring the meteoroid's compression or shear strength. This is the same dynamical pressure (0.9 – 5 MPa) at which ordinary chondrite meteoroids break during the second fragmentation phase and final disruption (Borovicka et al., 2020). Surviving meteorites from the Murchison CM2 chondrite had a shear strength for 2.5 cm-square cubes of 7.1 – 16.6 MPa (Ciceri et al., 2021). For Aguas Zarcas to disrupt only when the dynamical pressure reached 5 MPa requires that this remaining part was a monolithic boulder that did not have significant cracks from collisions. Indeed, carbonaceous chondrites do not show the bimodal distribution of dynamic pressures shown by ordinary chondrites, suggesting that cracks likely play a less important role during atmospheric entry (Borovicka et al., 2020).

Another indication that Aguas Zarcas was relatively strong material, is the irregular shape and size of many of the larger meteorites found on the ground. Shortly after the 25-km shattering, there was not a lot of additional fragmentation before slowing down to dark flight, resulting in deep regmaglypts.

The known fall location of meteorites on the ground is diagnostic for the fragmentation behavior. The predicted fall location for meteorites of 10 g, 100 g and 1 kg falling from 25 km altitude are shown as open triangles in Fig. 6. We used the wind sonde data measured at station 78583 MZBZ in Belize (Phillip Goldston International Airport) at 0h UTC on 24 April 2019. Most find locations are East or West of those calculated positions. The dispersion along the meteoroid track can result from different origin altitudes, different ablation behavior after fragmentation, aerodynamic lift, or significant forward/backward ejection velocities during the flare. Differences in wind drift due to meteorite shape would mostly cause a north-south (perpendicular) dispersion. Given most surviving mass likely came from the surviving fragment at 25 km, the concentration of finds ahead of and behind the central location might measure radial velocities during the final breakup (Jenniskens et al., 2022), suggesting that fragments were ejected forwards and backwards. This could be the case if the final piece was elongated and spinning along an axis perpendicular to the vertical plane of the trajectory.

## Cosmic Ray Exposure age

A diameter of ~60 cm defines a shielding depth, refining the CRE age from noble gas data (Table 7). The inferred CRE age from $^{21}$Ne (2.1 ± 0.7 Ma) is consistent with the 2.0 Ma CRE age derived from cosmogenic radionuclides, and combined results suggest the most likely CRE age is 2.0 ± 0.2 Ma.

Other CM2-type carbonaceous chondrites have ages in the range of 0.05–10 Ma, but with about 40% of the ages of <1 Ma (Nishiizumi & Caffee, 2012; Zolensky et al., 2021; Krietsch et al., 2021). The CRE age of Aguas Zarcas overlaps with a broad peak between 1.5–3.0 Ma labeled as "group 4" in Zolensky *et al.* (2021).

## Meteoroid origin in the asteroid belt

Aguas Zarcas adds a fifth orbit to a small sample of meteorites with measured orbits that have CM affinity according to oxygen and chromium isotopes. Those are listed in Table 10. These meteorites are all breccias, but have a range in lithologies, oxygen isotopes, alteration and heating. Some are classified as C (Sutter's Mill) or C1 ungrouped (Flensburg) on account of complicated breccias. Aguas Zarcas itself contains C1 and CM lithologies. An orbit derived for CM type specimen Murchison was only based on visual observations and constrained the perihelion distance close to Earth orbit (Halliday & McIntosh, 1990).

So far, the observed CM group meteoroids do not move on the low-eccentricity orbits of evolved C-class and B-class asteroids like Ryugu and Bennu, so they are not the product of the disintegration of such asteroids.

Two of the orbits had evolved to low perihelion distances (Table 10). The loss of helium as evident from the lower CRE age derived from $^{3}$He in Table 7, suggests that the orbit of Aguas Zarcas did come closer to the Sun in the past than the perihelion distance q = 0.983 au observed now. Prior heating may have resulted in loss of more fragile parts of this meteoroid such that only the stronger parts survived, now penetrating deep during entry in Earth's atmopshere.

Including Aguas Zarcas, four out of five CM-affinity meteorites arrived on a low inclination ($i < 3°$) orbit, while C1-ungrouped meteorite Flensburg arrived on a higher $i \sim 7°$ orbit (Fig. 13). The low inclination and $q \sim 1$ au orbit of Aguas Zarcas (and Murchison according to the orbit by Halliday & McIntosh, 1990) has a high impact probability with Earth (Morbidelli & Gladman, 1998). In the case of ordinary chondrites, which have much higher CRE-ages, this geometric effect does not show clearly in the observed orbits. They show a much wider range of inclinations (Jenniskens, 2024). This implies that CM chondrite meteoroids do not survive long enough for the terrestrial planets to increase their inclination distribution and the source of CM chondrites is at a low inclination in the asteroid belt.

The short CRE ages, and the eccentric and low-inclined orbits imply that many CM meteoroids do not survive the inner solar system environment very long due to processes other than collisions. When they break, they tend to catastrophically disrupt into much smaller meteoroids. The CRE ages of CM-types are shorter than the collisional lifetime of 30-100 cm meteoroids in the main asteroid belt, where the collision probability is highest. Possible fragmentation mechanisms include thermal stresses and YORP spin-up, instead (Jenniskens *et al.,* 2012). This process is also seen for the larger NEO, where especially darker NEO tend to disappear from the population over time as well (Granvik *et al.,* 2016; Morbidelli *et al.,* 2020).

So far, there is no correlation between CRE age and delivery resonance or orbital elements of the impacting orbit (Table 10). Maribo and Winchcombe had an orbit with semi-major axis of a = 2.50 AU near the 3:1 mean-motion resonance with Jupiter (Fig. 13). This suggests they were delivered to Earth via the 3:1 mean-motion resonance relatively quickly, before the resonance pumped up the inclination. Flensburg arrived from the 5:2 mean motion resonance at 2.82 AU and impacted on a relatively high inclination of 7° (Fig. 13) orbit. Sutter's Mill and Aguas Zarcas results are consistent with an origin in either the 3:1 or 5:2 mean motion resonance. If these CM chondrites have a common asteroid family source region, then that family is likely in a low inclined orbit near these resonances (Jenniskens *et al.,* 2012). Alternatively, if the source of these CM chondrites is a background population of small asteroids from the C-class population, then the density of objects is expected to be higher at low inclinations near the 3:1 and 5:2 resonances compared to the $\nu 6$ (Morbidelli & Gladman, 1998).

Figure 13 plots some of the possible source asteroid families. Perhaps the most likely source are recent collisions among the rich Themis family of asteroids at low i ~ 1.08° inclination in the outer main belt (a = 3.134 AU). The family shows signs of hydrated silicates (3-micron band of OH) and aqueous alteration with reflectance spectra similar to CM chondrites (Florczak *et al.*, 1999). That would explain delivery of CM chondrites from both the 5:2 and 3:1 resonances. Delivery to Earth via the 2:1 resonance is less likely due to perturbations by Jupiter at aphelion.

The young ~8 Ma old Veritas family in the outer asteroid belt (Fig. 13), with spectra similar to Themis, has also been suggested as the source of CM carbonaceous chondrites (e.g., Krietsch *et al.*, 2021; Brož et al., 2024). It is less likely that Aguas Zarcas, Sutter's Mill, Winchcombe and Maribo originated from Veritas, but perhaps ungrouped C1 type Flensburg originated from the Veritas asteroid family (Jenniskens, 2024). Flensburg arrived with an extremely short CRE age of

<0.01 Ma (Bischoff et al., 2021), giving little time to pump up the inclination from a low Themis-family like value.

Brož et al. (2024) identified the König family, a small ~51 Ma young asteroid family in the central Main Belt with a = 2.57 AU, i = 8.8º, as a possible contributing source of CM chondrites. Again, that inclination is higher than suggested by the current sample of CM-affinity meteorite orbits.

Previously, the large Polana family in the inner main belt was suggested as well, but it lacks hydrated bands in asteroid reflectance spectra that are mostly F-class (Tatsumi et al., 2022). It is now thought that Polana may be the source of Ryugu and Bennu, asteroids of CI affinity (Brož et al., 2024). The small (~300 asteroids) Sulamitis family of C-type asteroids with a = 2.463 AU and i = 5.04º in the inner main belt (Fig. 13) has been proposed as a source also (Jenniskens, 2020), but it has a higher inclination than most CM-type orbits and spectra similar to the Polana family (Arredondo et al., 2021).

## CONCLUSIONS

The CM2-chondrite Aguas Zarcas arrived on a steep 81.2 ± 0.6° inclined trajectory with a slow 14.6 ± 0.6 km/s entry speed. It initially had a ~60 cm diameter and a mass of about 250 kg. Half of the mass survived to at least ~32-km altitude. The final surviving part catastrophically fragmented at ~25 km altitude, when the dynamic pressure was about 5 MPa. Aguas Zarcas was a monolith with relatively few cracks and other weaknesses. Thanks to this strength and the resulting deep penetration depth, followed by a significant disruption, about 11% of the mass survived to the ground and was collected.

Unlike other CM affinity chondrites such as Sutter's Mill and Flensburg, which had CRE ages of <0.1 Ma, Aguas Zarcas avoided fragmentation in the interplanetary medium for 2.0 ± 0.7 Ma, perhaps due to its strength. In general, CM chondrites are detected when they are freshly delivered to the inner solar system by the resonance, they do not typically survive much longer than 2 Ma.

The impact orbit was low-inclined, with the semi-major axis suggesting delivery via the 3:1 or 5:2 mean-motion resonances. The large and low-inclined Themis family is a likely source asteroid family for CM chondrites.

*Acknowledgments* – The stones of Aguas Zarcas provided to Arizona State University for classification and curation were purchased from local finders and provided to ASU by Michael Farmer. We thank SETI Institute's Brenda Simmons for encouragement.

*Conflicts of interest statement* - This work was in part supported by the NCCR "Planet S" of the Swiss NSF. This work was also supported by NASA grant NNX14-AR92G (PJ), NNX16AD34G (QZY), 80NSSC18K0854 (PJ, KW, QZY) and by the NASA Ames Asteroid Threat Assessment Program (PJ).

**Table 1a.** Available videos of the Aguas Zarcas meteor with known camera locations. Listed are the geographic location, video framerate, the time offset from GLM, the systematic errors in Right Ascension and Declination, and the source.

| **Camera** | **Lat. (°N)** | **Long. (°W)** | **Alt. (m)** | **Fps (Hz)** | **Size** | **Δt (s)** | **Δα cosδ (°)** | **Δδ (°)** | **From:** | **Source:** |
|---|---|---|---|---|---|---|---|---|---|---|
| Turrialba Volcano | 10.01926 | 83.75655 | 3298 | 12 | 1920 x 1080 | -3.36 | +0.3 | +0.8 | RSN-UCR | Camera # 3 (black and white) |
| Heredia Gate | (10.00000)† | (84.12449) | (1138) | 12.5 | 640 x 360 | -.- | +2.3 | -1.8 | Columbia Digital | Gate Security camera |
| Quepos | 9.42669 | 84.16690 | 12 | 4 | 933 x 480 | +0.40 | -0.1 | -0.4 | Marina Pez Vela | Security camera |
| Turrialba | 9.91190 | 83.68013 | 652 | 60 | 1440 x836 | +91.57 | -0.1 | +0.9 | Eduardo Hernández Pereira | Liceo Experimental Bilingüe Turrialba |
| San Ramón | 10.08114* | 84.46955 | 1057 | 60 | 1280 x 720 | +206.10 | +2.9 | +2.0 | Oscar Mario Alvarado Vásquez | Dash Cam YI, 1080P60 |
| Heredia Dash | (10.00000)†* | (84.12449) | (1138) | 30 | 1280 x 720 | -.- | +0.7 | -0.2 | David Avendaño | Dash cam |
| San José | 9.93881 | 84.05504 | 1216 | 30 | 460 x 240 | -.- | -1.8 | +1.6 | Cámaras Viales | Rotonda de la Bandera |

Notes: *) Location of car dashcam during main flare; †) Adopted location at geographic center of Heredia.

**Table 1b.** Extracted astrometric data from individual cameras. Time of frame 1 is given, as well as the frame rate for those frames that were measured. "F" marks the position of the flare.

| | **Tur. Vulcano 2** | **Her. Gate** | **Quepos** | **Turrialba** | **San Ramón** | **Her. Dash** | **San José** |
|---|---|---|---|---|---|---|---|
| Start | 03:07:24.352 | 03:07:23.339 | 03:07:24.159 | 03:07:23.279 | 03:07:24.746 | 03:07:23.479 | 030724.179 |
| FPS | 12 | 12.5 | 8 | 20 | 30 | 10 | 10 |
| # | R.A. Dec. | R.A. Dec. | R.A. Dec. | R.A. Dec. | R.A. Dec. | R.A. Dec. | R.A. Dec. |
| 1 | 98.25 34.40 | 131.14 51.42 | 108.50 77.38 | 106.70 36.74 | 204.41 62.45F | 129.45 52.22 | 114.93 53.99 |
| 2 | 97.44 34.23 | 130.23 51.73 | 106.74 77.74 | 106.64 36.79 | 205.02 62.65 | 128.22 52.82 | 113.65 54.43 |
| 3 | 97.19 34.20 | 129.34 52.20 | 103.69 77.92 | 105.77 36.78 | 205.51 63.04 | 127.03 53.36 | 112.61 54.50 |
| 4 | 96.65 34.27 | 128.49 52.72 | 101.87 78.06 | 105.39 36.81 | 206.04 63.14 | 125.73 53.83 | 111.68 54.91 |
| 5 | 96.22 34.22F | 127.51 52.87 | 99.70 78.19F | 105.08 36.84 | 206.52 63.25 | 124.64 54.25 | 110.41 55.09 |
| 6 | 95.94 34.18 | 126.41 53.45 | 98.01 78.58 | 104.77 36.87 | 206.93 63.46 | 123.50 54.75 | 109.02 55.36F |
| 7 | 95.51 34.13 | 125.42 53.75 | 95.74 78.39 | 104.24 36.92 | 207.34 63.59 | 122.52 55.06 | 107.89 55.65 |
| 8 | 95.15 34.15 | 124.60 54.23 | 93.92 78.53 | 103.77 36.96 | 207.69 63.82 | 121.26 55.48 | 107.12 55.78 |
| 9 | 94.65 34.15 | 123.63 54.42 | 91.26 78.67 | 103.23 37.01 | 208.25 63.74 | 119.86 56.01 | 105.83 56.06 |
| 10 | 94.31 34.10 | 122.66 54.76 | 90.24 78.70 | 103.00 37.03 | 208.74 63.84 | 118.13 56.39 | 104.65 56.22 |
| 11 | 94.04 34.01 | 121.73 55.17 | 88.89 78.79 | 102.69 36.89 | 209.21 63.99 | 116.87 56.72 | 103.73 56.42 |
| 12 | 93.88 34.12 | 120.72 55.50 | 86.80 78.81 | 102.31 37.08 | 210.07 64.37 | 115.59 57.02 | 102.81 56.70 |
| 13 | 93.60 34.06 | 119.88 55.71 | 85.41 78.82 | 101.74 36.98 | 210.37 64.75 | 114.28 57.31F | 102.26 56.67 |
| 14 | 93.23 34.01 | 118.96 56.16 | -.- | 101.36 37.13 | 210.60 64.85 | 112.85 57.76 | 101.33 56.93 |
| 15 | 93.01 34.05 | 117.60 56.47 | -.- | 100.76 37.17 | 211.39 65.13 | 111.81 57.93 | 100.64 56.90 |
| 16 | 92.56 33.97 | 116.57 56.84 | -.- | 100.37 37.19 | 212.00 65.06 | 110.64 58.26 | 99.96 56.95 |
| 17 | 92.22 33.94 | 115.30 57.20 | -.- | 99.98 37.15 | 212.51 65.34 | 109.14 58.51 | 99.55 56.98 |
| 18 | -.- | 114.26 57.30F | | 99.67 37.10 | 212.99 65.45 | 107.75 58.92 | -.- |
| 19 | -.- | 113.27 57.56 | -.- | 99.44 37.11 | 213.40 65.72 | 106.41 59.08 | -.- |
| 20 | -.- | 111.95 57.79 | -.- | 99.28 37.12 | 213.93 65.86 | 105.50 59.28 | -.- |
| 21 | -.- | 110.92 58.04 | -.- | 98.99 37.13 | 214.50 65.94 | 104.61 59.53 | -.- |
| 22 | -.- | 109.89 58.27 | -.- | 98.27 37.15 | 214.88 66.04 | 103.75 59.63 | -.- |
| 23 | -.- | 108.79 58.43 | -.- | 97.96 37.16 | 215.21 66.11 | 103.04 59.76 | -.- |
| 24 | -.- | 108.15 58.60 | -.- | 97.65 37.17 | 215.75 66.29 | 102.32 59.89 | -.- |
| 25 | -.- | 107.46 58.79 | -.- | 97.26 37.11 | 216.09 66.55 | -.- | -.- |
| 26 | -.- | 106.66 58.87 | -.- | 96.87 37.18 | 216.34 66.70 | -.- | -.- |
| 27 | -.- | 105.82 58.97 | -.- | 96.32 37.19 | 216.87 66.80 | -.- | -.- |
| 28 | -.- | 105.26 59.03 | -.- | 95.85 37.19 | 217.37 67.00 | -.- | -.- |
| 29 | -.- | 104.59 59.18 | -.- | 95.62 37.19 | 217.71 67.12 | -.- | -.- |
| 30 | -.- | 103.88 59.26 | -.- | 95.23 37.19F | 218.09 67.32 | -.- | -.- |
| 31 | -.- | 103.45 59.30 | -.- | 95.07 37.19 | 218.94 67.50 | -.- | -.- |
| 32 | -.- | 103.06 59.41 | -.- | 94.68 37.19 | 219.35 67.64 | -.- | -.- |
| 33 | -.- | 102.66 59.52 | -.- | 94.21 37.25 | 219.65 67.72 | -.- | -.- |
| 34 | -.- | -.- | -.- | 93.97 37.25 | 220.09 67.81 | -.- | -.- |
| 35 | -.- | -.- | -.- | 93.66 37.24 | 220.52 67.95 | -.- | -.- |
| 36 | -.- | -.- | -.- | 93.42 37.17 | 220.91 67.95 | -.- | -.- |
| 37 | -.- | -.- | -.- | 93.19 37.17 | -.- | -.- | -.- |
| 38 | -.- | -.- | -.- | 92.97 37.16 | -.- | -.- | -.- |
| 39 | -.- | -.- | -.- | 92.71 37.22 | -.- | -.- | -.- |

**Table 2.** Trajectory and orbit of Aguas Zarcas at entry on April 24, 2019 UTC.

| | **This work** | **[1]** |
|---|---|---|
| *Trajectory (apparent):* | | |
| Date | 2019-04-24 | 2019-04-24 |
| Time Begin (UT) | 03:07:20.63 | -.- |
| Right Ascension (°, app.) | 165.8 ± 0.6 | -.- |
| Declination (°, apparent) | +13.2 ± 1.0 | -.- |
| Entry Speed (km/s, app.) | 14.6 ± 0.6 | 14 |
| Az of radiant from S (°) | 109.4 ± 5.8 (WNW) | 117 |
| El of radiant (°) | 81.2 ± 0.6 | 73 |
| Convergence Angle (°) | 77.9 | -.- |
| Altitude Begin Lightcurve (km) | 70.4 | -.- |
| | | |
| Altitude First Observed (km) | 42.0 | -.- |
| Latitude First Observed (°, N) † | 10.4164 | 10.4777 |
| Longitude First Observed (°, W) | 84.3915 | 84.5168 |
| | | |
| Time Flare (UT) | 03:07:24.779 | -.- |
| Altitude Flare (km) | 25.1 ± 1.0 | -.- |
| Latitude Flare (°, N) | 10.4092 | -.- |
| Longitude Flare(°, W) | 84.3710 | -.- |
| | | |
| Altitude End (km) | 18.2 | -.- |
| Latitude End (°, N) | 10.4054 | 10.4146 |
| Longitude End (°, W) | 84.3598 | 84.3905 |
| | | |
| *Orbit:* | | |
| Solar Longitude (°, J2000) | 33.4087 | -.- |
| Right Ascension (°, geoc.) | 161.7 ± 1.2 | -.- |
| Declination (°, geocentric) | +13.7 ± 8.3 | -.- |
| Entry Speed (km/s, geoc.) | 9.5 ± 1.0 | -.- |
| Perihelion Distance (AU) | 0.983 ± 0.006 | 0.999 |
| Semi-major Axis (AU) | 2.62 (+0.70/-0.46) | 2.7 |
| Eccentricity | 0.625 ± 0.080 | 0.63 |
| Inclination (°) | 1.40 ± 1.39 | 3.09 |
| Argument of Perihelion (°) | 199.4 ± 2.9 | 185.3 |
| Node (°) | 33.48 ± 0.40 | 33.4 |

Notes: [1] From Soto *et al.* (2019b), based on Turrialba Volcano and Quepos views only; †) Beginning point is that of beginning of record, not beginning of meteor.

**Table 3.** Aguas Zarcas known find location of meteorites (developed from Lücke *et al.,* 2019).

| # | Mass (g) | Lat. (°N) | Long. (°W) | Alt. (m) | Date of find (local time)† | Finder | Notes |
|---|---|---|---|---|---|---|---|
| 1 | 1152 | 10.39131 | 84.34125 | 363 | 2019-04-23 | Marcia Campos Muñoz | Pierced roof over hang in La Caporal |
| 2 | 1875* | 10.40665 | 84.35734 | 456 | 2019-04-23 | Family mother | On muddy field on farm [1] |
| 3 | 361 | 10.40181 | 84.36531 | 318 | 2019-04-24 | Two nieces of Víctor Julio Vargas Hernández | Grassy area next to road, 6cm deep crater (impact to az 310º) |
| 4 | 622 | 10.40115 | 84.36193 | 330 | 2019-04-24 | Farm worker | On rural concrete street |
| 5 | 280 | 10.40253 | 84.36425 | 321 | 2019-04-24 | Esmeralda | Pierced roof of dog house ("Rocky") [2] |
| 6 | 39.6 | 10.40630 | 84.35751 | 430 | 2019-04-25 | As #2, + daughter | Flat disk |
| 7a | 0.586 | 10.40674 | 84.35748 | 455 | 2019-04-25 | As #2, + daughter | |
| 7b | 0.298 | 10.40674 | 84.35748 | 455 | 2019-04-25 | As #2, + daugther | |
| 7c | 0.484 | 10.40674 | 84.35748 | 455 | 2019-04-25 | As #2, + daughter | |
| 7d | 0.998 | 10.40673 | 84.35740 | 455 | 2019-04-27 | As #2, + daughter | Recovered after rain |
| 8 | 24.1 | 10.40228 | 84.36014 | 342 | -.- | Oscar Murillo Salas | Recovered in Palmerita creek [3] |
| 9 | 13.3 | 10.40169 | 84.36418 | 323 | -.- | Geovanny Rojas Fonseca | Recovered in La Cocaleca [3] |
| 10 | -.- | 10.38524 | 84.33618 | 401 | -.- | -.- | -.- |
| 11 | -.- | 10.39352 | 84.33352 | 343 | -.- | -.- | -.- |
| 12 | -.- | 10.39826 | 84.33892 | 315 | -.- | -.- | -.- |
| 13 | -.- | 10.39843 | 84.35233 | 385 | -.- | -.- | -.- |
| 14 | -.- | 10.40199 | 84.35472 | 419 | -.- | -.- | -.- |
| 15 | -.- | 10.40427 | 84.35695 | 432 | -.- | -.- | -.- |
| 16 | -.- | 10.39936 | 84.35874 | 340 | -.- | -.- | -.- |
| 17 | -.- | 10.39196 | 84.35090 | 362 | -.- | -.- | -.- |
| 18 | -.- | 10.40521 | 84.36371 | 327 | -.- | -.- | -.- |
| 19 | -.- | 10.40791 | 84.36531 | 344 | -.- | -.- | -.- |
| 20 | -.- | 10.40814 | 84.36748 | 329 | -.- | -.- | -.- |
| 21 | -.- | 10.41118 | 84.37089 | 344 | -.- | -.- | -.- |
| 22 | -.- | 10.41020 | 84.37440 | 309 | -.- | -.- | -.- |
| 23 | -.- | 10.41174 | 84.37571 | 305 | -.- | -.- | -.- |
| 24 | -.- | 10.40223 | 84.36530 | 318 | -.- | -.- | -.- |
| 25 | -.- | 10.40018 | 84.36540 | 329 | -.- | -.- | -.- |
| 26 | -.- | 10.40133 | 84.36687 | 318 | -.- | -.- | -.- |
| 27 | -.- | 10.39843 | 84.36978 | 321 | -.- | -.- | -.- |
| 28 | -.- | 10.39600 | 84.37004 | 332 | -.- | -.- | -.- |
| 29 | -.- | 10.39253 | 84.37159 | 342 | -.- | -.- | -.- |
| 30 | -.- | 10.40022 | 84.37135 | 314 | -.- | -.- | -.- |
| 31 | -.- | 10.40154 | 84.37358 | 305 | -.- | -.- | -.- |
| 32 | -.- | 10.40133 | 84.37905 | 293 | -.- | -.- | -.- |
| 33 | 14 | 10.40591 | 84.35678 | 460 | -.- | Brayan Arraya | Find documented [4] |

Notes: † Rain hit the area starting on afternoon of April 27, 2019, lasting three days; *) dry weight; [1]) Now at Field Museum in Chicago; [2]) Now stored at ASU; [3]) Now at National Museum of Costa Rica; [4]) From youtube video.

**Table 4.** Oxygen isotopes of Aguas Zarcas.

| Sample | Sub-Sample | mg | date | $\delta^{18}O'$ | $\delta^{17}O'$ | $\Delta^{17}O'$ | Src: |
|---|---|---|---|---|---|---|---|
| ASU#2121_5 | 2a | 1.7 | 29-Aug-19 | 14.223 | 4.662 | -2.848 | This |
| | 2c | 2.8 | 29-Aug-19 | 13.796 | 3.621 | -3.663 | This |
| | 2b | 2.1 | 29-Aug-19 | 13.642 | 3.518 | -3.685 | This |
| | 2f | 1.6 | 29-Aug-19 | 12.626 | 3.327 | -3.340 | This |
| | 2d | 2.5 | 29-Aug-19 | 11.533 | 2.186 | -3.903 | This |
| | 2e | 1.7 | 29-Aug-19 | 11.107 | 1.505 | -4.360 | This |
| Pre-rain | 1f | 2.7 | 3-May-19 | 12.669 | 3.959 | -2.730 | [1] |
| | 1d | 2.3 | 3-May-19 | 12.314 | 3.843 | -2.659 | [1] |
| | 1e | 2.1 | 3-May-19 | 12.249 | 3.995 | -2.473 | [1] |
| | 1c | 2.6 | 3-May-19 | 11.591 | 3.360 | -2.760 | [1] |
| | 1b | 2.5 | 3-May-19 | 10.807 | 2.932 | -2.774 | [1] |
| | 1a | 2.7 | 3-May-19 | 10.342 | 2.381 | -3.080 | [1] |
| | 1g | 2.8 | 3-May-19 | 9.748 | 2.143 | -3.004 | [1] |
| Block | 2a | few | -.- | 7.39 | 0.98 | -2.922 | [2] |
| | 2b | few | -.- | 7.06 | 0.76 | -2.968 | [2] |

Notes: $\Delta^{17}O'$ linearized with slope 0.528; [1] Soto *et al.* (2019b); [2] Findlay *et al.* (2020).

**Table 5.** Bulk chromium isotope compositions of Aguas Zarcas and other carbonaceous chondrites with CM affinity for which entry orbits were measured.

| Sample | $\varepsilon^{53}Cr$ | $\varepsilon^{54}Cr$ | Source |
|---|---|---|---|
| Aguas Zarcas ASU:#2121_5 | +0.23 ± 0.06 | +1.09 ±0.11 | this |
| Aguas Zarcas A (less altered) | -.- | +0.98 ± 0.13 | [1] |
| Aguas Zarcas B (more altered) | -.- | +0.88 ± 0.13 | [1] |
| Flensburg (Zürich) | +0.20 ± 0.04 | +0.95 ± 0.05 | [2] |
| Flensburg (Münster) | +0.19 ± 0.05 | +1.06 ± 0.11 | [2] |
| Murchison | -.- | +1.10 ± 0.15 | [1] |
| Murchison | +0.12 ± 0.04 | +0.89 ± 0.09 | [3] |
| Murchison | +0.16 ± 0.04 | +0.89 ± 0.08 | [4] |
| Sutter's Mill SM 43 | +0.14 ± 0.04 | +0.95 ± 0.09 | [4] |
| Sutter's Mill SM 51 | +0.12 ± 0.04 | +0.88 ± 0.07 | [4] |
| Winchcombe | +0.32 ± 0.03 | +0.78 ± 0.07 | [5] |

Notes: [1] Torrano *et al.* (2021); [2] Bischoff *et al.* (2021); [3] Yin *et al.* (2009, 2010); [4] Jenniskens *et al.* (2012), Yamakawa & Yin (2014); [5] Greenwood *et al.* (2024).

**Table 6a.** Measured He, Ne, Ar concentrations (in $10^{-8}$ $cm^3/g$) as well as concentrations of trapped $^{20}Ne$ and $^{36}Ar$ (in $10^{-8}$ $cm^3/g$) and isotopic ratios ($^3He/^4He$ × 10 000).

| Sample | Mass | $^4He$ | $^3He/^4He$ | $^{20}Ne$ | $^{20}Ne_{tr}$ | $^{20}Ne/^{22}Ne$ | $^{21}Ne/^{22}Ne$ |
|---|---|---|---|---|---|---|---|
| Large | 28.446±0.022 | 3438±27 | 2.973±0.031 | 13.92±0.12 | 13.61±0.15 | 6.235±0.032 | 0.2206±0.0015 |
| Small | 17.017±0.012 | 3446±28 | 3.093±0.033 | 13.57±0.16 | 13.07±0.17 | 5.704±0.041 | 0.2055±0.0016 |

| Sample | Mass | $^{36}Ar$ | $^{36}Ar/^{38}Ar$ | $^{40}Ar/^{36}Ar$ |
|---|---|---|---|---|
| Large | 28.446±0.022 | 73.73±0.29 | 5.299±0.025 | 18.22±0.30 |
| Small | 17.017±0.012 | 73.11±0.33 | 5.271±0.046 | 14.00±0.37 |

Notes: Uncertainties (1σ) include ion counting statistics, mass discrimination, and blank variations. Concentrations additionally include uncertainty on sensitivity variations and $^{20}Ne_{tr}$ includes uncertainty introduced by the deconvolution.

**Table 6b.** Measured concentrations of $^{84}Kr$ (in $10^{-10}$ $cm^3/g$) and Kr isotopic ratios (×100).

| Sample | $^{84}Kr$ | $^{78}Kr/^{84}Kr$ | $^{80}Kr/^{84}Kr$ | $^{82}Kr/^{84}Kr$ | $^{83}Kr/^{84}Kr$ | $^{86}Kr/^{84}Kr$ |
|---|---|---|---|---|---|---|
| Large | 96.73±0.45 | 0.5995±0.0069 | 3.896±0.024 | 19.985±0.094 | 20.018±0.094 | 30.75±0.14 |
| Small | 94.50±0.54 | 0.6010±0.0066 | 3.891±0.033 | 20.05±0.15 | 19.99±0.15 | 30.96±0.22 |

Notes: Uncertainties (1σ) include ion counting statistics, mass discrimination, blank variations, and for $^{84}Kr$ sensitivity variations.

**Table 6c.** Measured concentrations of $^{132}Xe$ (in $10^{-10}$ $cm^3/g$) and Xe isotopic ratios (×100).

| Sample | $^{132}Xe$ | $^{124}Xe/^{132}Xe$ | $^{126}Xe/^{132}Xe$ | $^{128}Xe/^{132}Xe$ | $^{129}Xe/^{132}Xe$ | $^{130}Xe/^{132}Xe$ | $^{131}Xe/^{132}Xe$ | $^{134}Xe/^{132}Xe$ | $^{136}Xe/^{132}Xe$ |
|---|---|---|---|---|---|---|---|---|---|
| Large | 106.25 | 0.4684 | 0.4082 | 8.168 | 105.75 | 16.130 | 81.87 | 37.98 | 31.91 |
| | ±0.75 | ±0.0028 | ±0.0036 | ±0.031 | ±0.42 | ±0.061 | ±0.31 | ±0.17 | ±0.13 |
| Small | 107.94 | 0.4657 | 0.4159 | 8.195 | 105.77 | 16.190 | 82.08 | 38.14 | 32.05 |
| | ±0.74 | ±0.0035 | ±0.0037 | ±0.041 | ±0.35 | ±0.053 | ±0.38 | ±0.15 | ±0.12 |

Notes: Uncertainties (1σ) include ion counting statistics, mass discrimination, blank variations, and for $^{132}Xe$ sensitivity variations.

**Table 7.** Concentrations of cosmogenic noble gases (in $10^{-8}$ $cm^3/g$), production rates (in $10^{-8}$ $cm^3/g/Ma$), and Cosmic Ray Exposure ages (in Ma).

| Sample | $^{21}Ne_{cos}$ | $^{3}He_{cos}$ | P(3) | P(21) | CRE(3) | CRE(21) |
|---|---|---|---|---|---|---|
| Large | 0.4425±0.0045 | 0.66±0.18 | 1.486–1.862 | 0.165–0.255 | 0.36–0.45 | 1.74–2.69 |
| Small | 0.4343±0.0053 | 0.71±0.18 | 1.486–1.862 | 0.165–0.255 | 0.38–0.48 | 1.71–2.64 |
| Combined | | | | | 0.41 ± 0.07 | 2.1 ± 0.7 |

Notes: Uncertainties (1σ) on concentrations include ion counting statistics, mass discrimination, blank and sensitivity variations as well as uncertainty introduced by the deconvolution.

**Table 8a.** Massic activities (corrected to date of fall of the meteorite April 24[rh], 2019) of cosmogenic radionuclides (in dpm $kg^{-1}$) in the specimens of the Aguas Zarcas stone measured by non-destructive gamma-ray spectroscopy. For comparison are also given data from the Maribo CM2 chondrite (Haack *et al.,* 2012). Errors include a 1σ uncertainty of 10% in the detector efficiency calibration.

| **Nuclide** | **Half-life** | **Maribo** (four specimens together) (9.95 g) | **Aguas Zarcas** Flight-oriented (2.645 g) | Pre-rain (3.2813 g) |
|---|---|---|---|---|
| $^{7}Be$ | 53.22 d | 238 ± 29 | < 220 | < 300 |
| $^{58}Co$ | 70.83 d | 8 ± 2 | 26 ± 8 | < 12 |
| $^{56}Co$ | 77.236 d | 5 ± 2 | < 19 | < 9.2 |
| $^{46}Sc$ | 83.787 d | 11 ± 2 | 20 ± 5 | 11 ± 4 |
| $^{57}Co$ | 271.8 d | 10 ± 2 | 15 ± 3 | 7 ± 2 |
| $^{54}Mn$ | 312.3 d | 95 ± 7 | 100.0 ± 6.4 | 56.9 ± 3.8 |
| $^{22}Na$ | 2.60 y | 59 ± 5 | 40.3 ± 3.8 | 44.5 ± 3.5 |
| $^{60}Co$ | 5.27 y | 10 ± 1 | 32 ± 3 | 2 ± 1 |
| $^{44}Ti$ | 60 y | 2.6 ± 0.5 | < 11 | < 3.6 |
| $^{26}Al$ | $7.17 \times 10^5$ y | 33 ± 3 | 32.0 ± 2.8 | 35.6 ± 2.7 |

**Table 8b.** Concentration of primordial radionuclides (ng $g^{-1}$ for U and Th chains and μg $g^{-1}$ for K) in the specimens of the Aguas Zarcas stone measured by non-destructive gamma-ray spectroscopy. For comparison are also given data from the Maribo CM2 chondrite (Haack *et al.,* 2012). Errors include a 1σ uncertainty of 10% in the detector efficiency calibration.

| **Element** | **Maribo** | **Aguas Zarcas** | |
|---|---|---|---|
| | (four specimens together) | Flight-oriented | Pre-rain |
| | (9.95 g) | (2.645 g) | (3.2813 g) |
| U | 16 ± 3 | 17 ± 4 | 14 ± 3 |
| Th | 54 ± 5 | 60 ± 10 | 50 ± 10 |
| K | 275 ± 24 | 210 ± 30 | 360 ± 30 |

**Table 9.** Measured concentrations of major/minor elements and cosmogenic radionuclides in Aguas Zarcas (CM) chondrite.

| **Element/ Nuclide** | **Aguas Zarcas** 104.1 mg | **Average CM** [1] |
|---|---|---|
| Mg (wt%) | 11.8 | 11.7 |
| Al (wt%) | 1.11 | 1.18 |
| P (wt%) | 0.11 | 0.09 |
| S (wt%) | 3.3 | 3.3 |
| K (ppm) | 530 | 400 |
| Ca (wt%) | 1.31 | 1.27 |
| Ti (ppm) | 590 | 580 |
| Mn (wt%) | 0.16 | 0.17 |
| Fe (wt%) | 21.4 | 21.0 |
| Co (ppm) | 580 | 575 |
| Ni (wt%) | 1.22 | 1.20 |
| $^{10}$Be (dpm/kg) | 15.50 ± 0.14 | 20–28* |
| $^{26}$Al (dpm/kg) | 39.3 ± 0.9 | 32–56* |
| $^{36}$Cl (dpm/kg) | 60.5 ± 1.2 | 6–8* |

* Expected $^{10}$Be, $^{26}$Al and $^{36}$Cl production rates for CM chondrites with R = 20–150 cm are from model calculations of Leya & Masarik (2009); [1] Wasson (2012).

**Table 10.** Classification criteria, collision history, and source region constraints of CM affinity chondrites with known pre-atmospheric orbits.

| | **Sutters Mill** | **Maribo** | **Flensburg** | **Winchcombe** | **Aguas Zarcas** |
|---|---|---|---|---|---|
| *Classification:* | *[1,2]* | *[3]* | *[8]* | *[13,14]* | *[10,11]* |
| Type | C (CM2) | CM2 | C1 ung. | CM2 | CM2 |
| Lithologies | Highly brecciated | Relatively Unbrecc. | Small sample | Brecciated CM2.0-2.6 | Highly brecciated |
| Aqueous alteration stage | Moderate | Low | -.- | Low+high | Low+high |
| Thermal alteration stage | Moderate | Low | -.- | Low | Low+high |
| Sulfides/metal | Sulfides | Sulfides | -.- | -.- | Metal |
| Chondrule size | -.- | Not well developed | ø 125 μm | ø 145 μm | ø 100–300 μm; 16 or 23 vol% |
| *Chronology:* | *[1,2]* | *[3]* | *[5,6,7]* | *[14]* | *This work* |
| CRE age (Ma) | ~0.019 | 0.9 ± 0.3 | 0.006 ± 0.001 | 0.3 | 2.1 ± 0.5 |
| Collisional lifetime (Ma)§ | 17.1 | 6.6 | 7.7 | 4.6 | 7.7 |
| K-Ar age (Ma) | ~4500 | -.- | -.- | -.- | -.- |
| Mn-Mn isochron age (Ma) | 4563.7 ± 1.3 | -.- | 4564.6 ± 1.0 | -.- | -.- |
| *Orbit:* | *[1]* | *[4]* | *[8,9]* | *[12]* | *This work* |
| R.A. geocentric (°, J2000) | 24.0 ± 1.3 | 125.0 ± 0.3 | 183.46 ± 0.11 | 56.64 ± 0.02 | 161.7 ± 1.2 |
| Decl. geocentric (°) | +12.7 ± 1.7 | +19.8 ± 0.2 | -18.18 ± 0.14 | 17.71 ± 0.07 | +13.7 ± 8.3 |
| V geocentric (km/s) | 26.0 ± 0.7 | 25.8 ± 0.3 | 15.97 ± 0.06 | 8.12 ± 0.13 | 9.5 ± 1.0 |
| Semi-major axis (AU) | 2.59 ± 0.35 | 2.43 ± 0.12 | 2.82 ± 0.03 | 2.586 ± 0.008 | 2.62 (+0.70/-0.46) |
| Eccentricity | 0.824 ± 0.020 | 0.805 ± 0.010 | 0.701 ± 0.003 | 0.618 ± 0.001 | 0.625 ± 0.080 |
| Perihelion distance (AU) | 0.456 ± 0.022 | 0.475 ± 0.005 | 0.843 ± 0.001 | 0.98684 ± 0.00001 | 0.983 ± 0.006 |
| Inclination (°) | 2.38 ± 1.16 | 0.25 ± 0.16 | 6.82 ± 0.06 | 0.46 ± 0.01 | 1.40 ± 1.39 |
| Argument of Perihelion (°) | 77.8 ± 3.2 | 279.4 ± 0.6 | 307.25 ± 0.16 | 351.80 ± 0.02 | 199.4 ± 2.9 |
| Node (°) | 32.77 ± 0.06 | 297.46 ± 0.15 | 349.207 ± 0.001 | 160.196 ± 0.001 | 33.48 ± 0.40 |

Notes: §) Collisional lifetime in the main belt is based on size and is taken as 1.4 $\sqrt{r}$, with r the radius in cm (Jenniskens, 2018). Source: [1] Jenniskens *et al.* (2012), Jilly *et al.* (2014), Nishiizumi *et al.* (2014); [2] Ott *et al.* (2013); [3] Haack *et al.* (2012); [4] Borovicka et al. (2019); [5] Roth *et al.* (2008); [6] Dominik & Jessberger (1979); [7] Bischoff *et al.* (2021); [8] Heinlein *et al.* (2020); [9] Borovicka *et al.* (2021); [10] Kouvatsis & Cartwright (2020); [11] Kerraouch *et al.* (2021); [12] McMullan *et al.* (2024); [13] Daly *et al.* (2024); [14] King *et al.* (2022).

**Fig. 1.** Location of video cameras relative to the meteor trajectory and flare (marked by a star) over the fall location near Aguas Zarcas, Costa Rica. Scale lower right is 70 km.

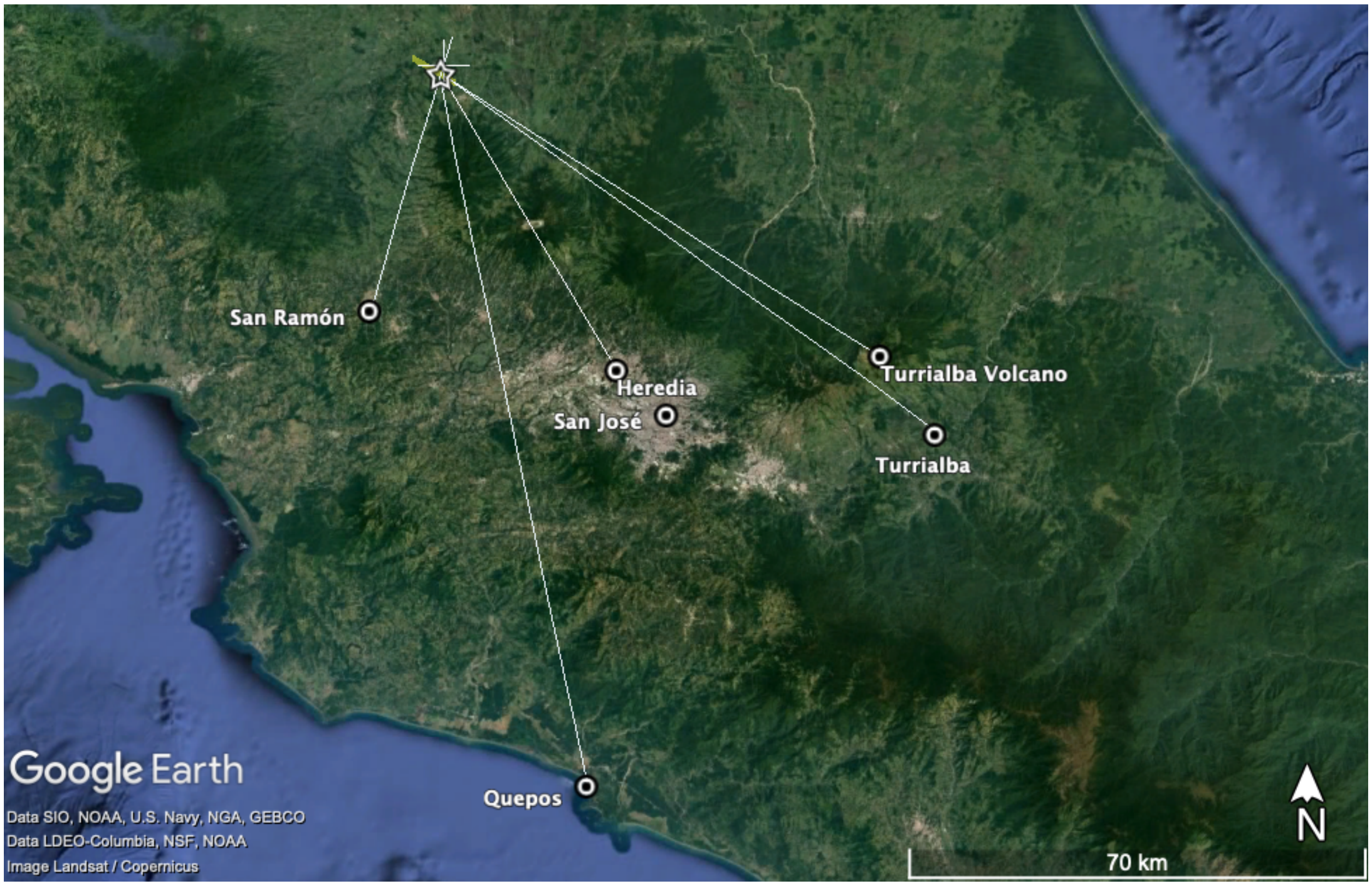

**Fig. 2.** Meteor videos analyzed here: **(A)** Single frame of *Turrialba Volcano 2* video just after the final flare. Stars Sirius (left) and Betelgeuse (center) are marked. Meteor in top right corner; **(B)** Turrialba video with meteor near center. Street lamps used for calibration are marked by azimuth directions from North; **(C)** Heredia security camera video with black points tracing the motion of the meteor; **(D)** Quepos video with meteor top left; **(E)** San Ramón video with meteor top center; **(F)** San José Rotonda de la Bandera security camera video with meteor top left. White lines in each video show the placement of the horizon.

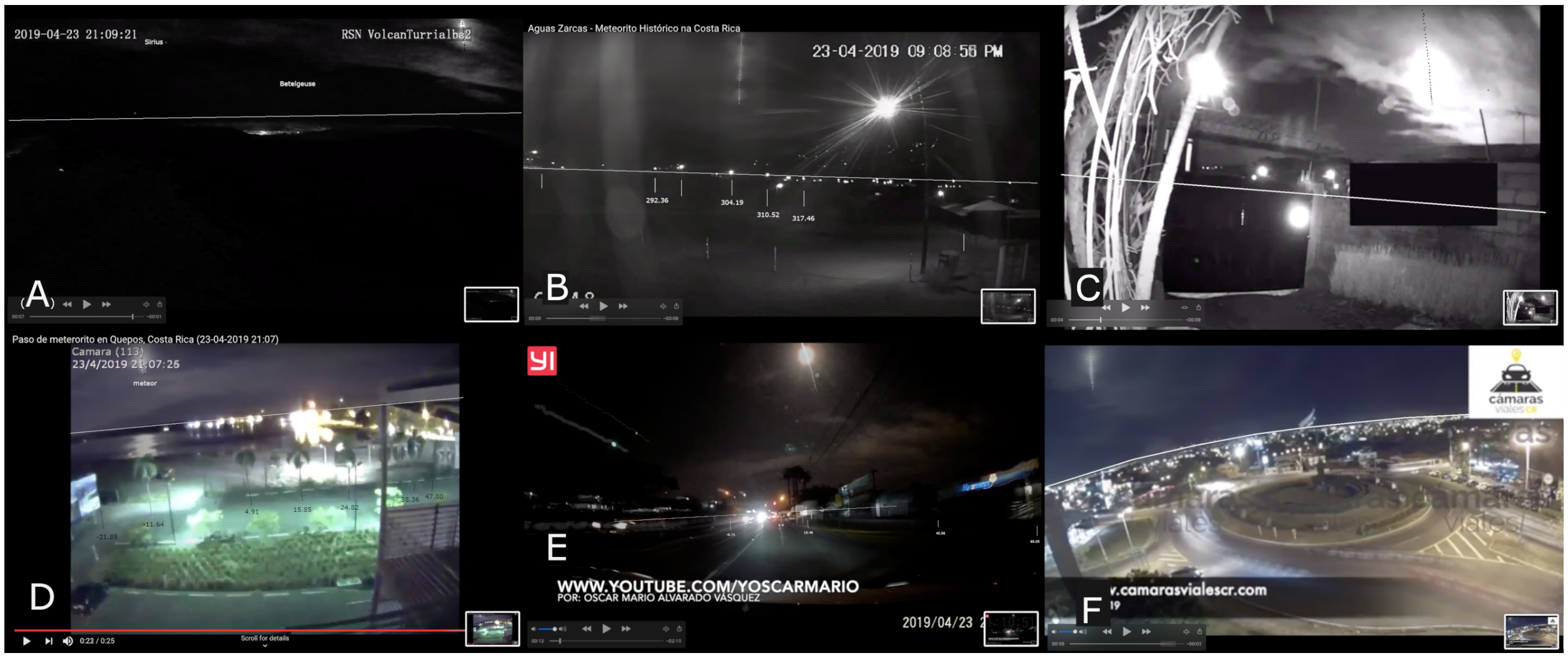

**Fig. 3.** Example of alt-azimuth grid overlay on the Heredia dashboard camera video. Perspective lines on the road define the position of the horizon, and the orientation of a lamp post in different video frames (white lines left, slightly bend halfway up) and features on buildings (right) defines the rotation to normal. The flare elevation (assuming an altitude of 25.1 km) defines the scale. In this case, the camera's central azimuth changed over time and the meteor path in the central frame shown was adjusted using the meteor wake as guidance.

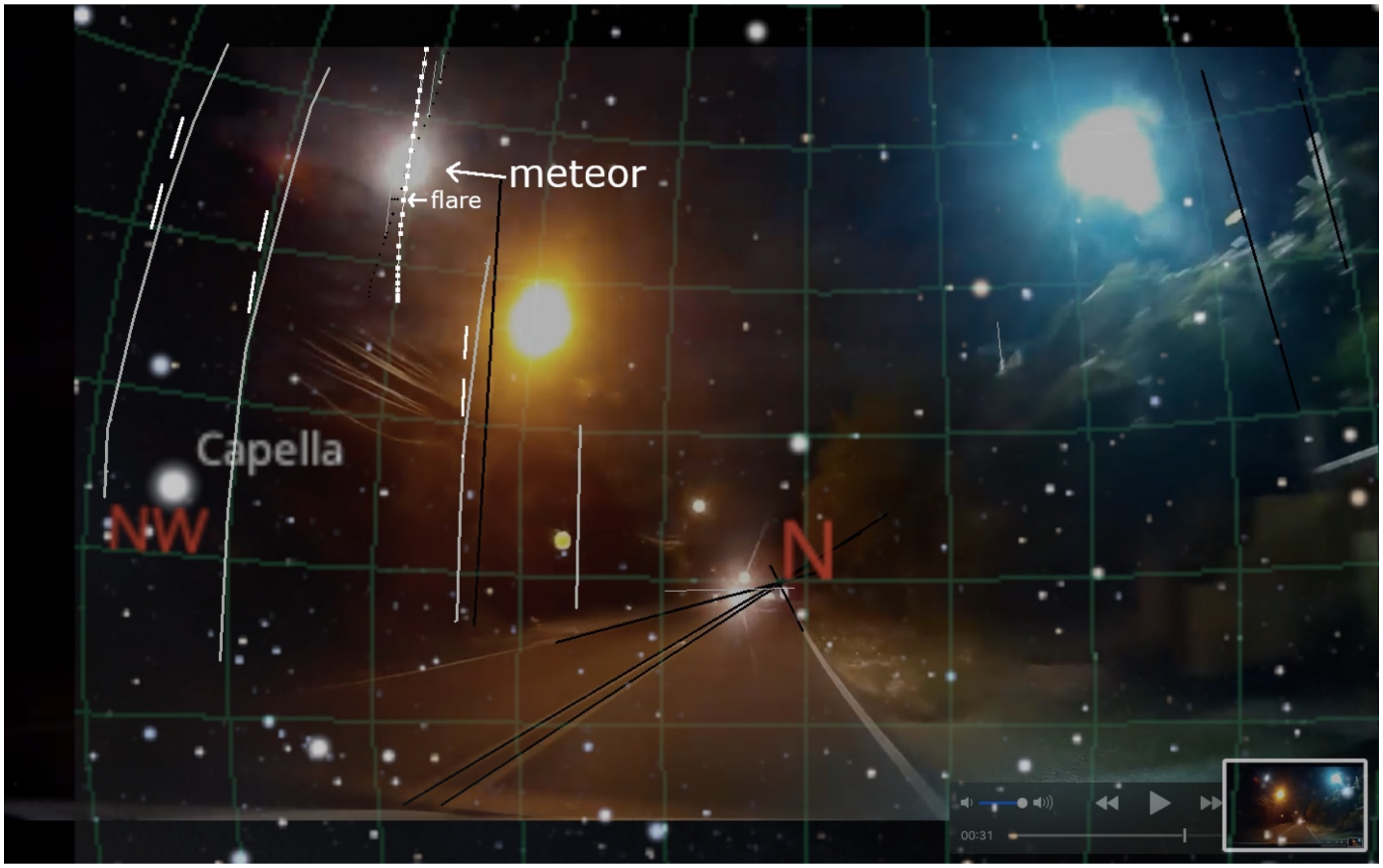

**Fig. 4.** Height versus time profile for all lines of sight, after aligning the altitude of the flare to 25.1 km and aligning the post-flare velocity profile of Heredia gate (HE) and Heredia dashboard camera (HED) to that of cameras with known locations. The dashed line fits the early part of the Turrialba video (TU) where the deceleration is still small.

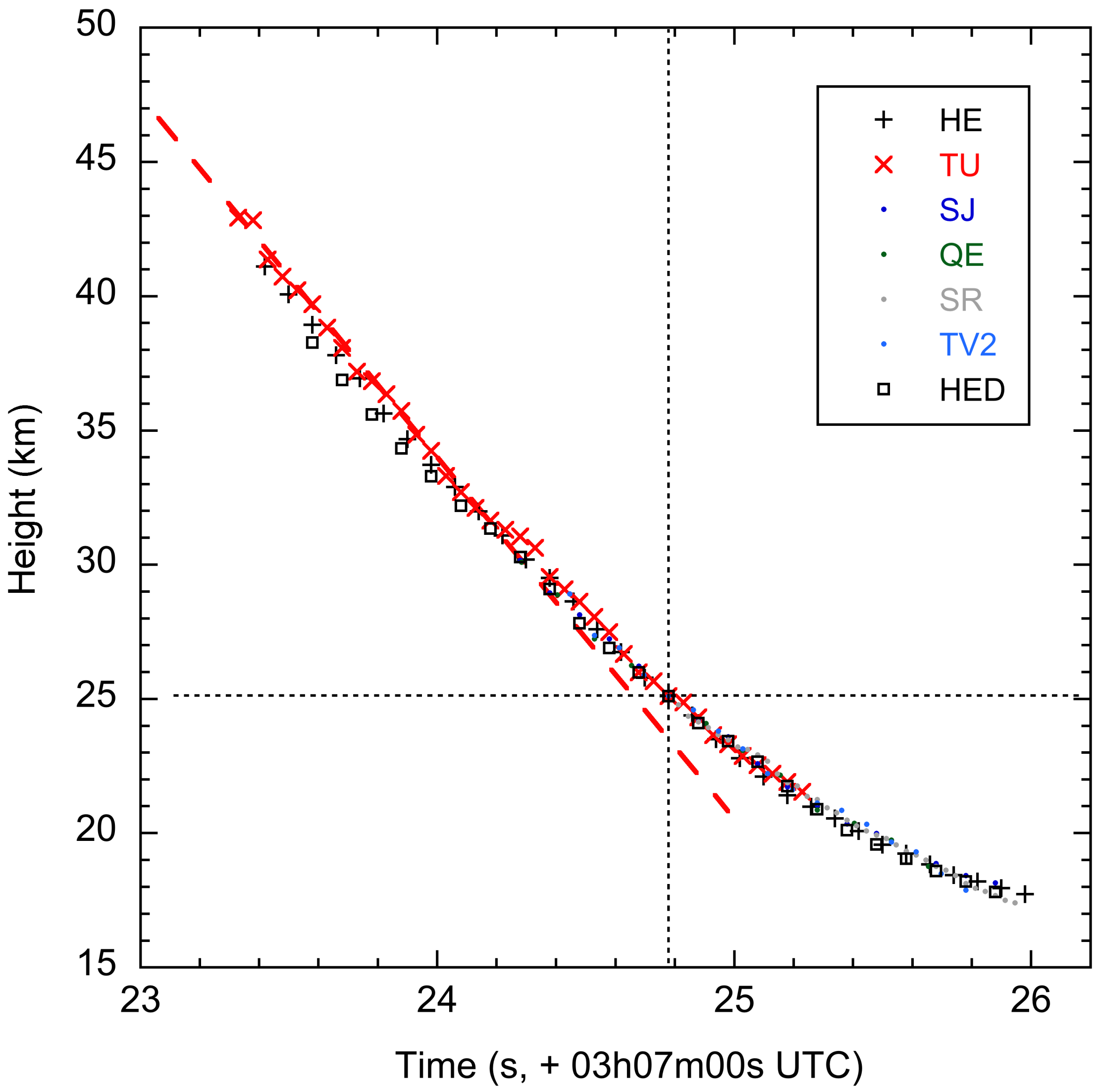

**Fig. 5.** Light curve of the Aguas Zarcas meteor, plotting the magnitude at 100 km distance versus time. Points lower than the contour are affected by atmospheric extinction, including high altitude clouds.

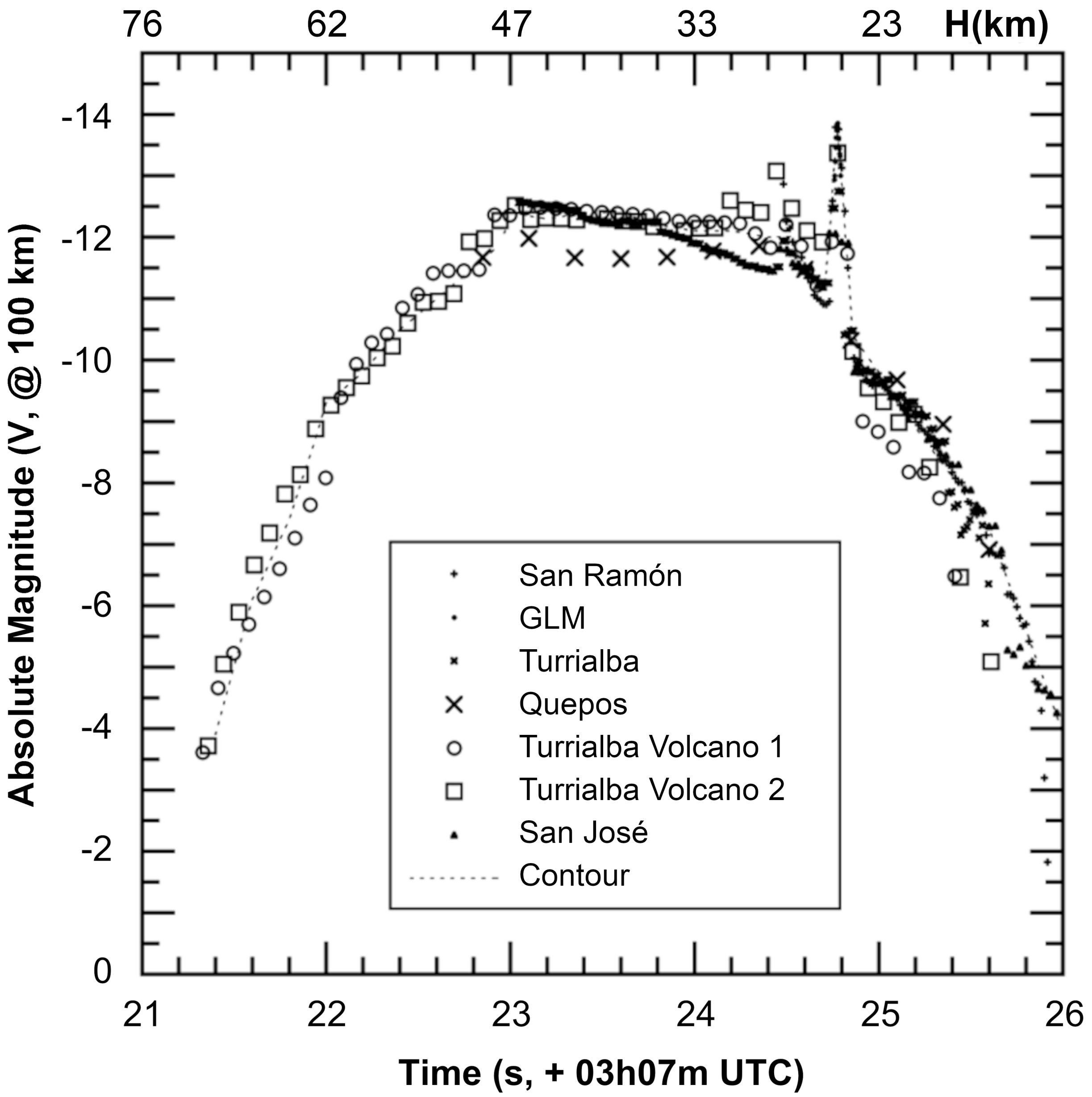

**Fig. 6.** Meteorite strewn field with meteor trajectory (moving highly inclined from left to right). The position of the meteoroid during the final flare is marked by a bullseye. The predicted fall location for masses of 10 g, 100 g, and 1 kg falling from this flare are marked by open triangles. Each meteorite's find location is labeled with the number listed in Table 3. Also marked is the measured GLM position of the flare, with error bars.

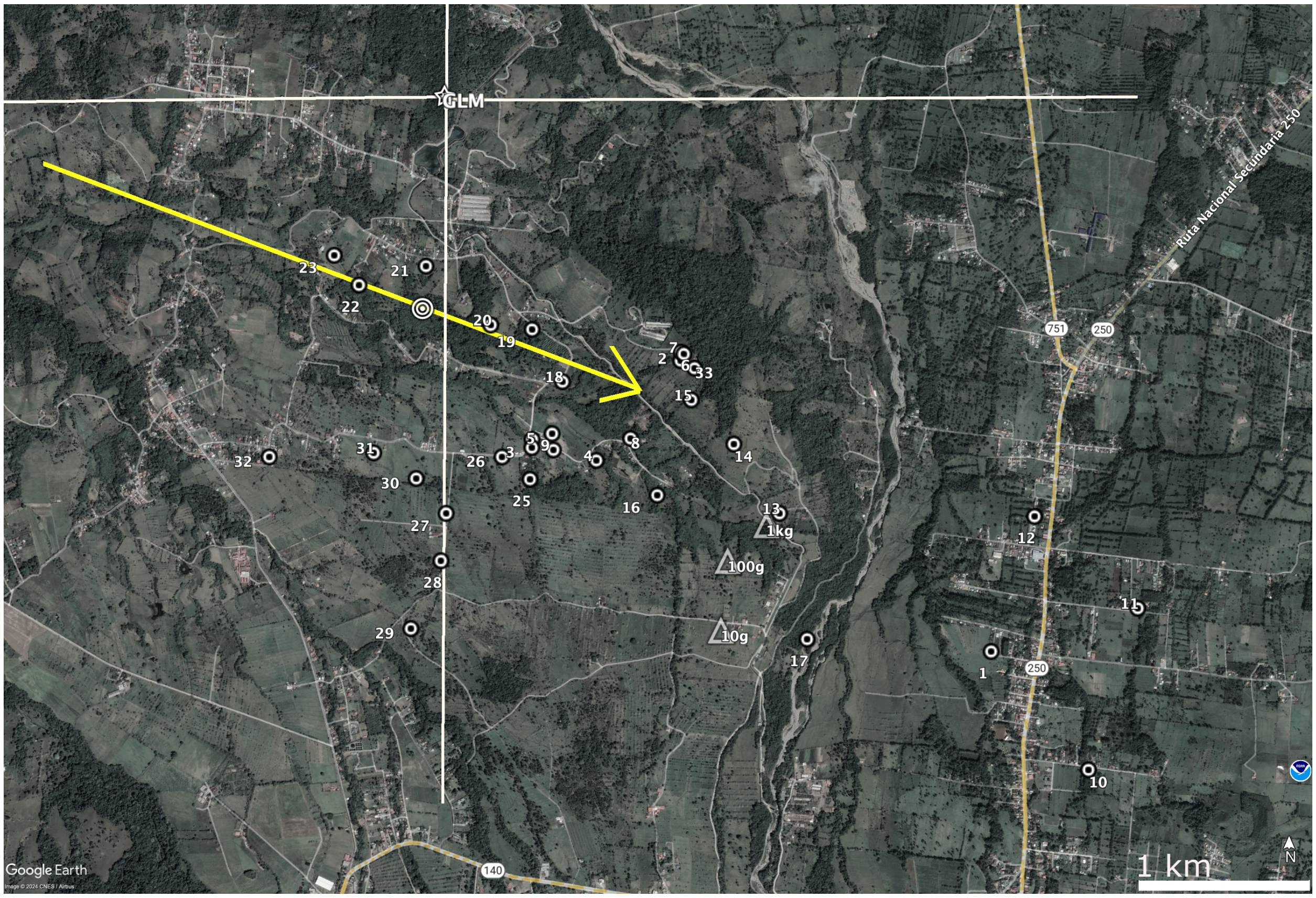

**Fig. 7.** Aguas Zarcas meteorite. This 146.2 g stone shows irregular surface features from ablation without the relatively flat surfaces that result from secondary fragmentation.

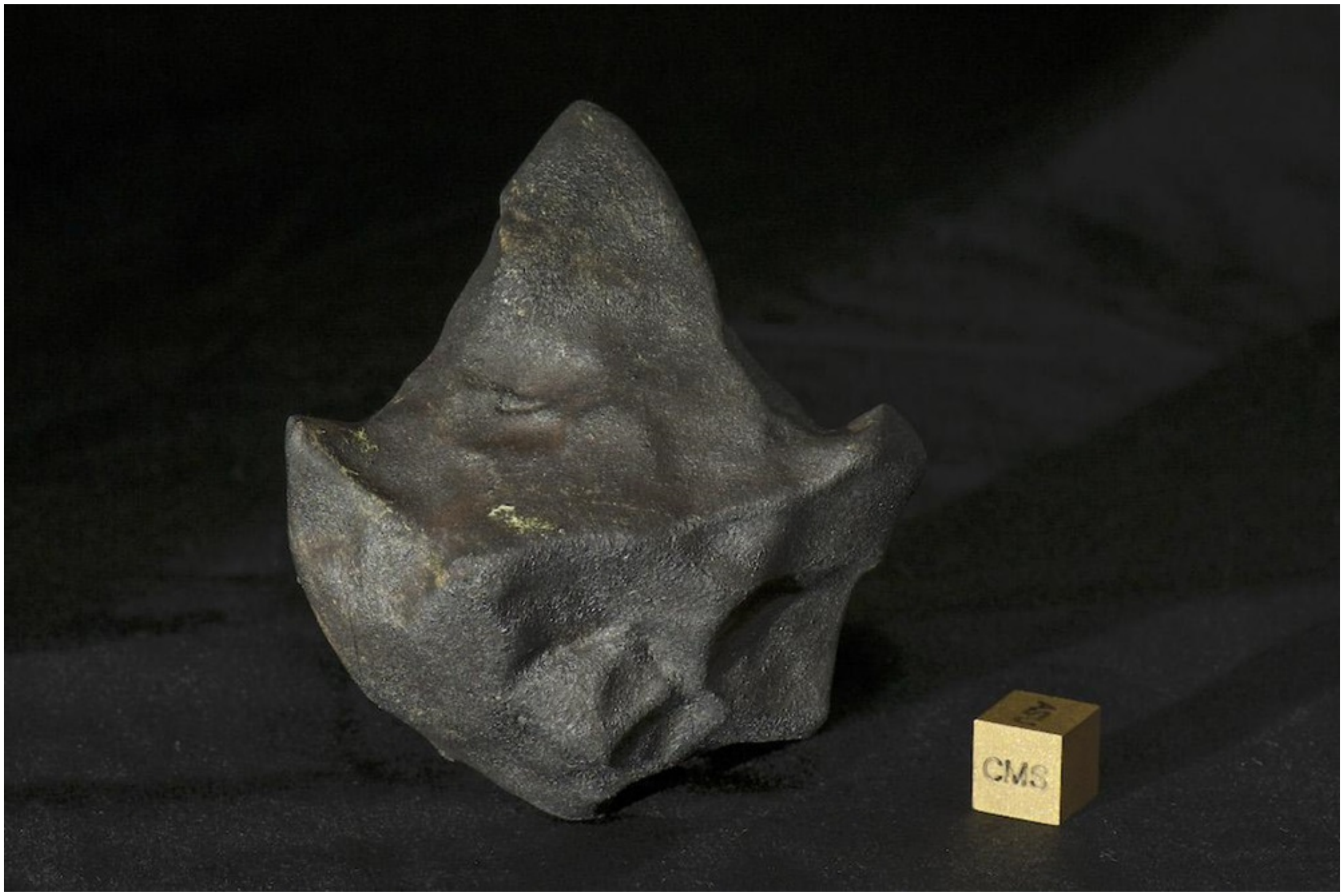

**Fig. 8.** Results from Cr and oxygen isotopic analyses of Aguas Zarcas.

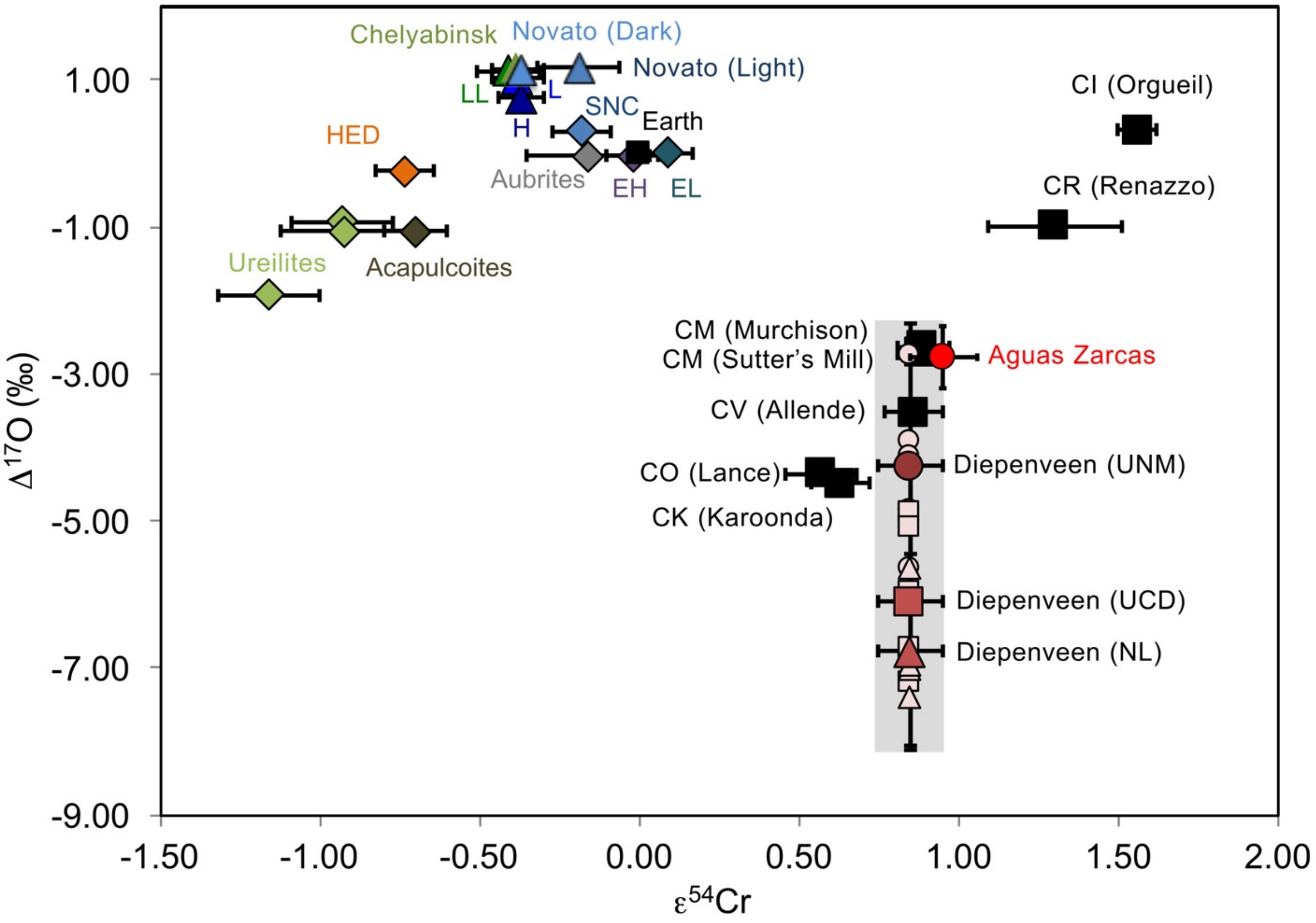

**Fig. 9.** Noble gas results. Ne three-isotope plot showing the isotopic composition of the two aliquots of Aguas Zarcas as well as endmembers commonly detected in CM chondrites. The Ne in Aguas Zarcas is a mix between trapped Ne with a $^{20}Ne/^{22}Ne \sim 7.0\text{-}7.8$ and cosmogenic Ne. The composition of the trapped Ne agrees with a mixture of the HL component carried by presolar nanodiamonds, and/or Q gases carried by Phase Q, and the Ne-E component known to be carried by presolar SiC and graphite grains. Components are marked Q (Busemann *et al.,* 2000), air (Ozima & Podosek, 2002), HL (Huss & Lewis, 1994), cosmogenic (cos) (Wieler, 2002), and Ne-E (Amari *et al.,* 1995; Lewis *et al.,* 1994).

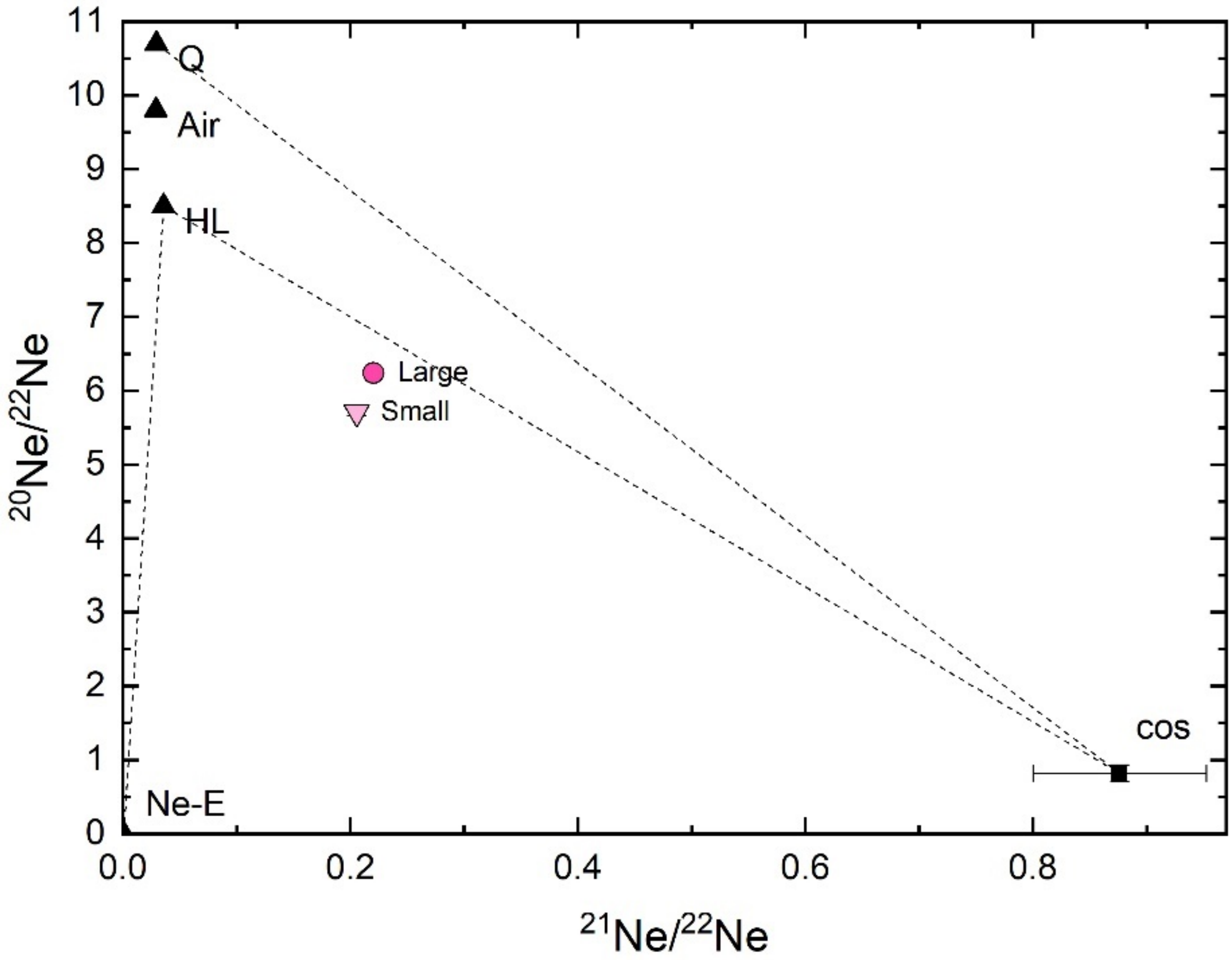

**Fig. 10.** Noble gas results. The Xe isotopic composition of the two aliquots of Aguas Zarcas normalized to $^{132}Xe$ and air. Xenon is dominated by the Q component with small amounts, ~2% of HL gases from presolar nanodiamonds. Components are marked Air (Basford *et al.,* 1973), Q (Busemann *et al.,* 2000), SW (Meshik *et al.,* 2014), and HL (Huss & Lewis, 1994; Ott, 2014).

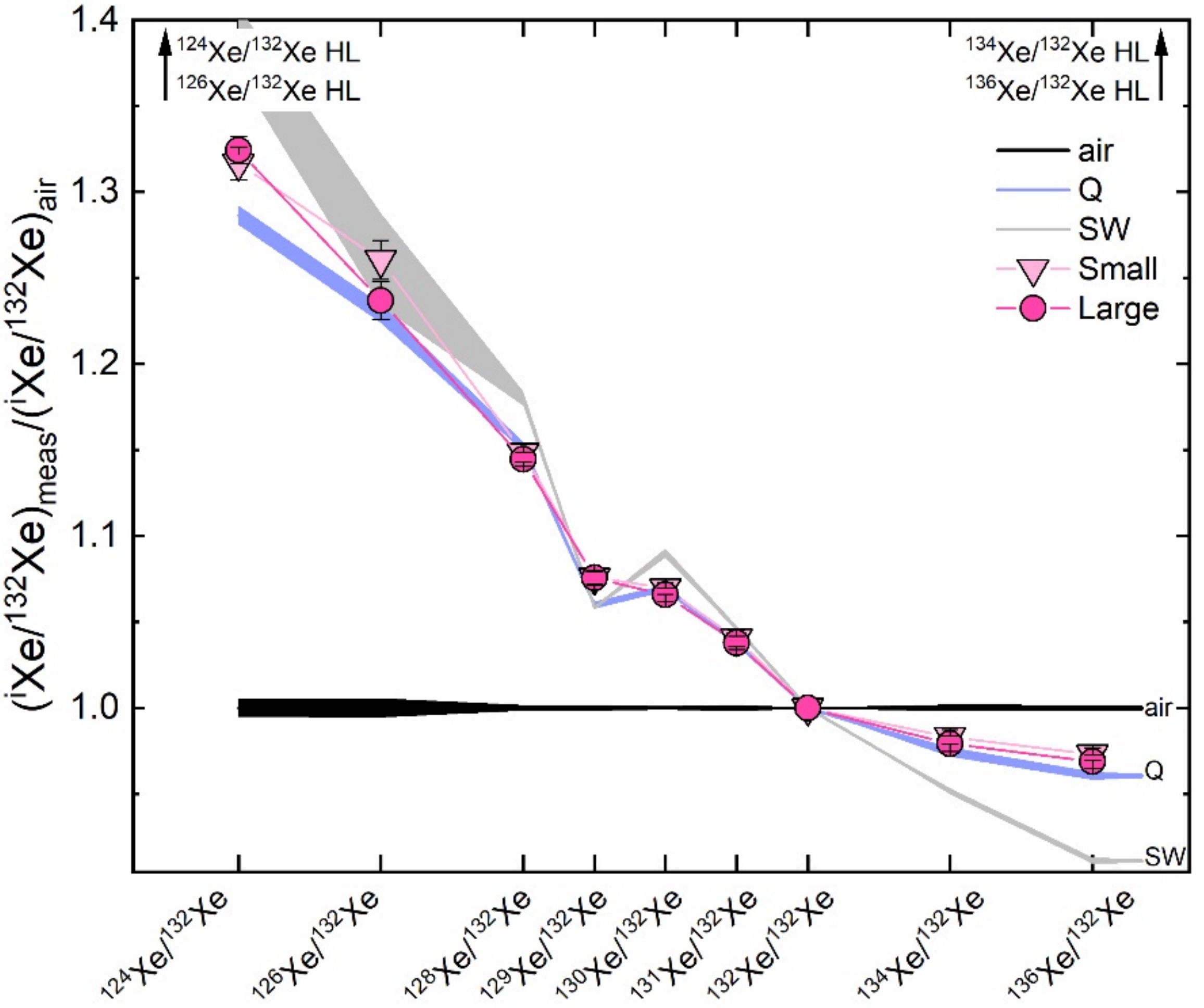

**Fig. 11.** Measured cosmogenic radionuclide concentrations in Aguas Zarcas as a function of depth in a spherical meteoroid in comparison with calculated production rates in CM chondrites (Leya & Masarik, 2009). The arrows in (a) and (b) represent the corrections for undersaturation of $^{10}$Be and $^{26}$Al assuming a CRE age of 2.0 Ma, which yields the best fit with calculated production rates, as shown by the dashed curve in panel (c), which represents corrections of $^{10}$Be and $^{26}$Al for CRE ages ranging from 1.5 Ma to 15 Ma. The measured $^{36}$Cl concentration of 60.5 dpm/kg in Aguas Zarcas is a factor of 7–10 higher than calculation production rates for spallation from K, Ca, Ti and Fe, indicating Aguas Zarcas contains a large contribution of $^{36}$Cl from neutron-capture on $^{35}$Cl.

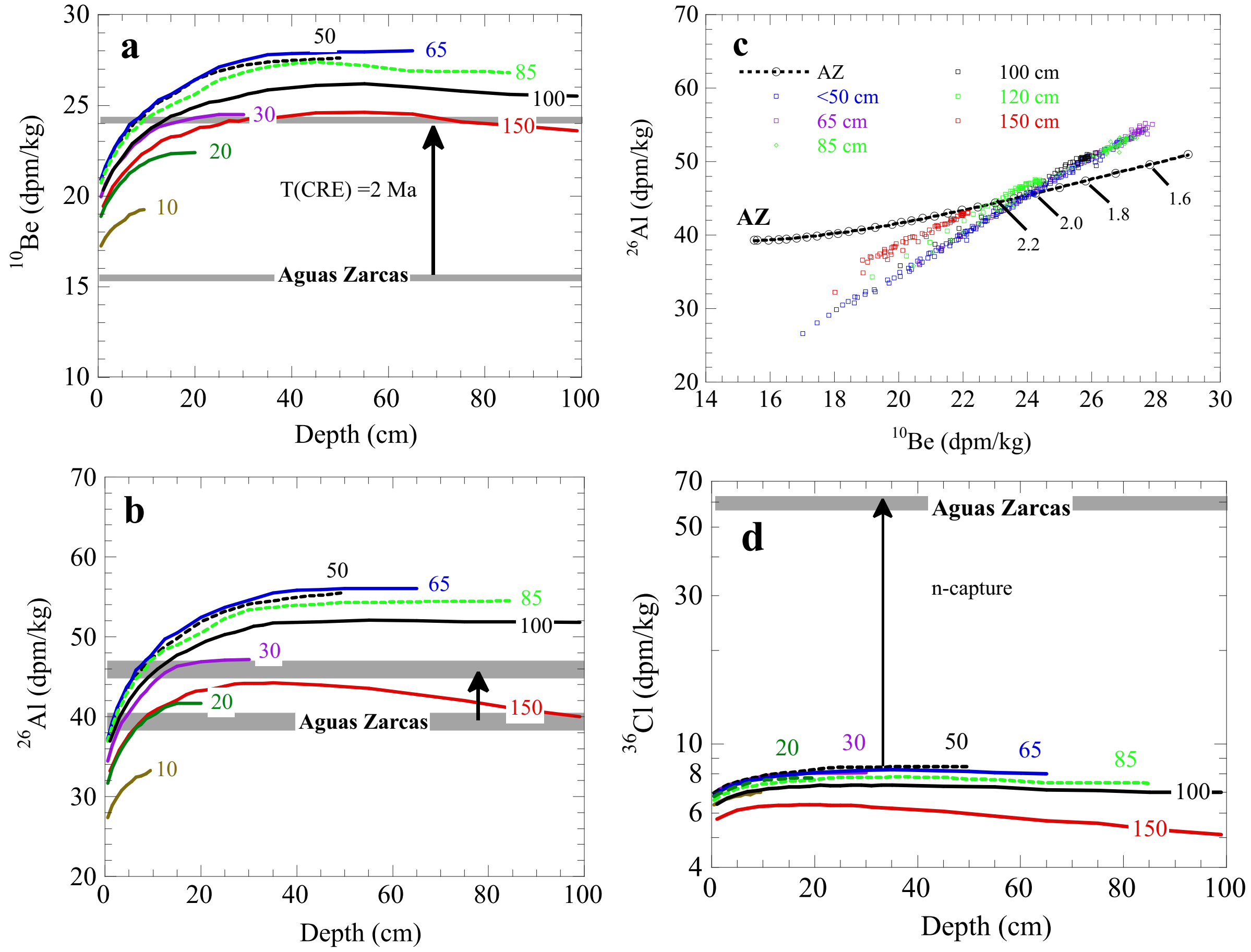

**Fig. 12.** Burst altitude versus yield strength for a 14.2 km/s entry speed at 81.2 degrees entry angle.

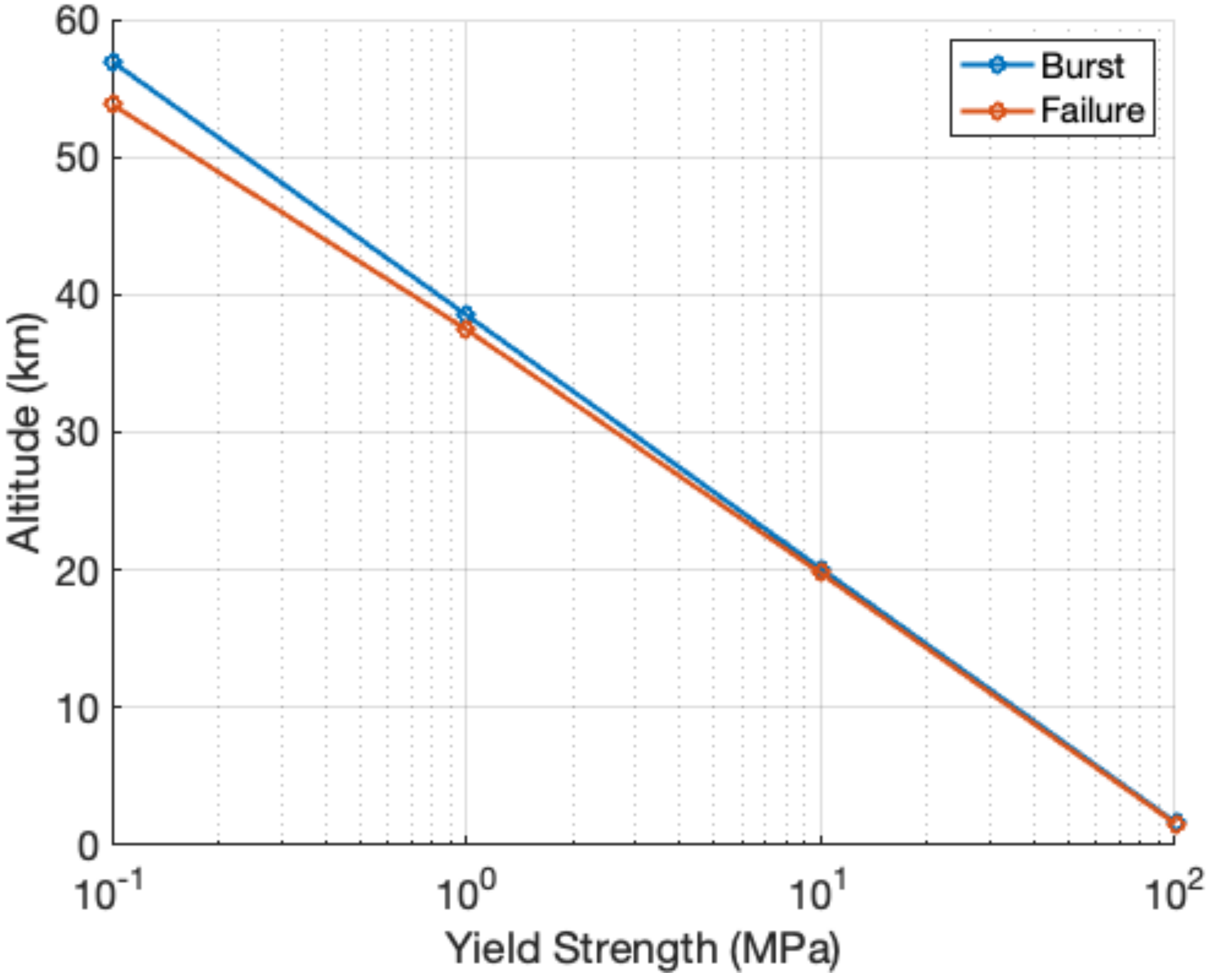

**Fig. 13.** Orbital elements of the impacting orbit for all observed meteorite falls of CM affinity. Dashed lines mark the delivery resonances. Potential source asteroid families are marked by a cross.

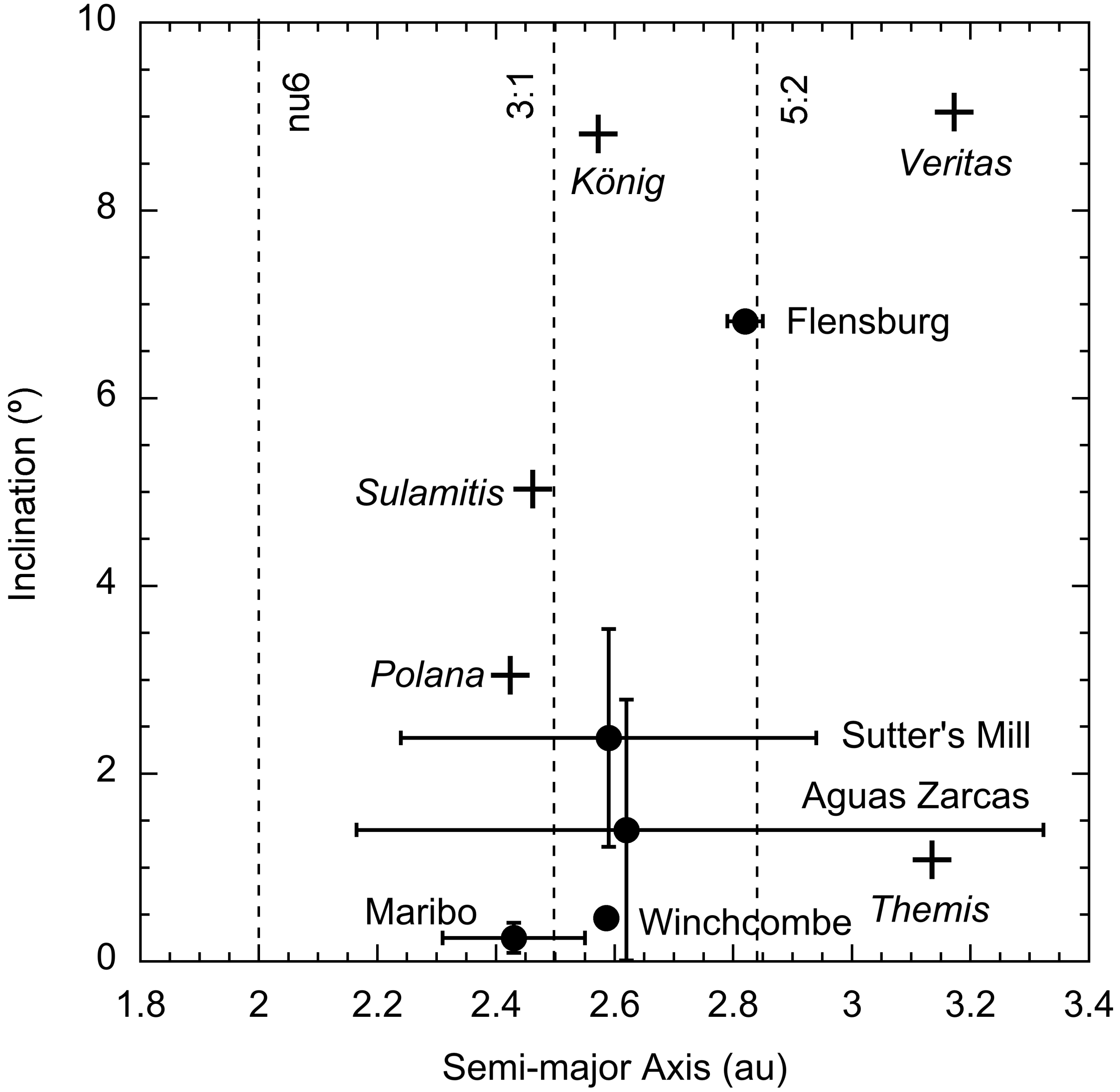